\documentclass[11pt,a4paper,english]{scrreprt}
\usepackage{ae}
\usepackage{aecompl}
\usepackage{amsmath}
\usepackage{graphicx}
\usepackage{amssymb}
\usepackage{latexsym}
\usepackage{multicol}
\usepackage{multirow}
\usepackage{babel}
\usepackage{acronym}
\usepackage{enumerate}
\usepackage[numbers]{natbib}
\usepackage{tabularx}
\usepackage[utf8]{inputenc}
\usepackage{makecell} 
\usepackage{adjustbox}
\usepackage{listings}
\usepackage{xcolor}
\usepackage{tcolorbox}
\usepackage{scrhack}
\usepackage{svg}
\usepackage{float}
\usepackage{csquotes}
\usepackage[utf8]{inputenc}

\lstdefinestyle{pythonstyle}{
    language=Python,                    
    basicstyle=\ttfamily\footnotesize,         
    keywordstyle=\color{blue}\bfseries, 
    stringstyle=\color{red},            
    commentstyle=\color{green!50!black},
    showstringspaces=false,             
    frame=lines,                        
    numbers=left,                       
    numberstyle=\tiny\color{gray},      
    stepnumber=1,                       
    tabsize=4,                          
    breaklines=true,                    
    breakatwhitespace=true,             
    backgroundcolor=\color{gray!10},    
    captionpos=b,                       
    morekeywords={self, None},          
}

\setuptoc{lof}{totoc}

\usepackage{tocbasic}
\addtotoclist[bibliography]{Bibliography}
\KOMAoptions{bibliography=totoc}

\typearea[current]{current}

\begin{document}


\titlehead{{Brandenburg Technical University Cottbus\\ Software Systems Engineering\\ Chair of Practical Computer Science}}

\subject{Master Thesis \\ \vspace{0.5cm}\includegraphics[width=0.5\textwidth]{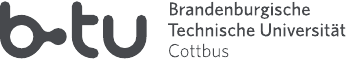}}

\title{Extracting Knowledge Graphs from User Stories Using Langchain}

    \author{Thayná Camargo da Silva \\ Matriculation Number: 5003023 \\ Master in Artificial Intelligence}

\date{\textit{29.07.2024} \\ \textit{13.01.2025}}

\publishers{

1. Examiner: Prof. Dr. rer. nat. Leen Lambers \\ 2. Examiner: Dr. Kate Cerqueira Revoredo, Humboldt-Universität zu Berlin 

}

\maketitle

\chapter*{Declaration}
The author declares that he / she has written the thesis at hand independently, without outside help and without the use of any other but the listed sources. Thoughts taken directly or indirectly from external sources (including electronic sources) are marked accordingly without exception. Sources used verbatim and contentual were quoted according to the recognised rules for scientific working. This thesis has not been submitted in the same or similar form, not even partially, within the scope of a different examination.

Thus far it also has not been publicised yet.

I herewith agree that the thesis will be examined for plagiarism with the help of a plagiarism-detection service.

\vspace{2cm} 
\noindent Place, Date: \rule{5cm}{0.4pt} \hfill Signature: \rule{5cm}{0.4pt}

\chapter*{Acknowledgments}
Completing this thesis represents an important professional and personal milestone for me, and  I am deeply grateful to those who supported me throughout this process.

I would first like to express my gratitude to Prof. Dr. Leen Lambers, who allowed me to be part of her research team. She not only recommended the initial inspiration for this research but also actively provided me with guidance and valuable feedback from day one. I would also like to extend my thanks to Dr. Kate Revoredo who accepted the invitation to collaborate on this research and brought a new perspective to it. Both are outstanding scientists and serve as an inspiration to me as I begin my journey in the technology domain.

My sincere appreciation goes to Dr. Sébastien Mosser for his contribution, sharing his valuable experience, providing crucial data, and offering insightful feedback that greatly contributed to this research. I am also grateful to the Artificial Intelligence study program at BTU, especially my mentor, Prof. Dr. Douglas Cunningham, for his support and guidance throughout my academic journey.

To my friends around the world, thank you for your encouragement during both the highs and the lows of this process. Finally, I am deeply grateful to my family, who prioritized education throughout my life and encouraged me to pursue my goals with determination.

\chapter*{Abstract}
Requirements engineering is a fundamental component in software engineering since it defines the software from the user's perspective. In this context, user stories play a crucial role since they are simple, natural language descriptions of software requirements, and they are widely utilized in the industry, along with the agile development methodology.

User stories, while valuable for capturing individual functionalities, offer a limited perspective of the overall system, hindering maintainability and comprehension. To address these challenges, extracting structured information and modeling user stories is crucial. Knowledge graphs offer a promising approach by providing a visual and structured representation of user stories, facilitating data storage and analysis, and reducing manual effort retrieval, leading to a more coherent and manageable system.

Several methodologies to model stories employ \ac{NLP} techniques, presenting limited precision, complex implementation, and difficulties to interpret a sentence~\cite{raharjana2021user}. Recent research has explored using large language models, such as ChatGPT-3.5 \cite{arulmohan2023extracting}, to extract knowledge graph components (nodes and relationships) from user stories. However, the solution relies on specific models and lacks comprehensive data processing pipelines for constructing knowledge graphs.

LangChain, a model-agnostic framework, is a promising tool that empowers the development of applications centered around large language models. Its adaptability extends to knowledge graph construction, facilitating the extraction of structured information from textual data and seamless integration with graph databases.

This thesis introduces a novel methodology for the automated generation of knowledge graphs from user stories by leveraging the advanced capabilities of \acp{LLM}. Utilizing the LangChain framework as a basis, the \acf{USGT} module was developed to extract nodes and relationships from user stories using an \ac{LLM} to construct accurate knowledge graphs.This innovative technique was implemented in a script to fully automate the knowledge graph extraction process. Additionally, the evaluation was automated through a dedicated evaluation script, utilizing an annotated dataset for assessment. By enhancing the visualization and understanding of user requirements and domain concepts, this method fosters better alignment between software functionalities and user expectations, ultimately contributing to more effective and user-centric software development processes.

\tableofcontents

\chapter*{Acronyms}
\begin{acronym}[USGT] 
\acro{AI}{Artificial Intelligence}
\acro{CRF}{Conditional Random Fields}
\acro{FN}{False Negative}
\acro{FP}{False Positive}
\acro{IE}{Information Extraction}
\acro{KG}{Knowledge Graph}
\acro{LCEL}{LangChain Expression Language}
\acro{LLM}{Large Language Model}
\acro{LSTM}{Long Short-Term Memory}
\acro{MC}{Multiple-Classification}
\acro{NER}{Named Entity Recognition}
\acro{NLP}{Natural Language Processing}
\acro{POS}{Part-of-Speech}
\acro{QA}{Question Answering}
\acro{RAG}{Retrieval-Augmented Generation}
\acro{RE}{Requirements Engineering}
\acro{RNN}{Recurrent Neural Network}
\acro{TP}{True Positive}
\acro{TS}{Token-Similarity}
\acro{USGT}{UserStoryGraphTransformer}
\end{acronym}

\chapter{Introduction}
\label{chap:introduction}
\section{Motivation}
Technology became a constant across all facets of society, influencing how companies operate, how education is delivered, and how public sectors function. In essence, technology permeates the various aspects of daily living, including economic growth. Extending the theory proposed by Clayton M. Christensen~\cite{christensen2016know}, which suggests that customers 'hire' products or services to accomplish specific tasks, it becomes integral to fully comprehend and address user expectations. This understanding is essential to delivering products that provide real value to users.

Central to the success of capturing user needs is the discipline of Requirements Engineering. \textit{\ac{RE}} \cite{sommerville2011software} is a fundamental domain which precedes the software development, that focuses on defining and managing the functionalities of software from the user's perspective.  A study by Robert Glass ~\cite{glass2002facts} concluded that most IT system errors occur because of wrong or missing requirements. Conversely, through effective \ac{RE} practices, teams can ensure that the software not only meets technical specifications but also aligns with the needs and expectations of its users.

Communicating requirements is a complex challenge. User stories \cite{cohn2004user}, semi-structured natural language texts, are a popular tool for capturing requirements, by simplifying complex requirements into concise, understandable segments that capture both the functionality desired by the user and the value it provides. However, in a complex application, user stories can be hard to manage, and might lead to development delays \cite{lucassen2017extracting}.

While user stories are a valuable tool for capturing requirements, modeling them using a \textit{\ac{KG}} offers significant advantages \cite{arulmohan2023extracting} \cite{lucassen2017extracting} \cite{ladeinde2023extracting}. A \ac{KG} provides a structured representation of the information contained in user stories, facilitating the identification of relationships between them, uncovering potential inconsistencies or redundancies, and enabling a deeper requirements analysis. In this context, the product backlog—a collection of N user stories—is represented as a unified knowledge graph, where each user story is modeled as a set of nodes, and their relationships are captured as edges. In essence, a \ac{KG} promotes the transition from a collection of user stories, also known as product backlog,  to a comprehensive understanding of user requirements. Therefore, this thesis proposes the creation of a knowledge graph to model user stories, aiming to enhance the clarity, coherence, and utility of the captured requirements.

In the age of \textit{\ac{AI}}, \textit{Large Language Models (LLMs)} \cite{chang2023survey} and \textit{Knowledge Graphs (KGs)} have become fundamental technologies, driving significant advancements in how machines understand, reason, and process natural language. LLMs have proven to have excellent ability to assist in a wide array of tasks, such as generating human-like text, and retrieving information to answer questions \cite{khorashadizadeh2024research}. Among these innovations, LangChain \cite{langchain-git} has emerged as a robust framework for developing LLM-powered applications. It provides an integrated solution for constructing knowledge graphs, greatly simplifying the knowledge extraction process, by reducing the complexity involved in interacting with \ac{LLM}s and offering modules that completely automate the creation of Knowledge Graphs. This thesis aims to explore the suitability of LangChain as a tool to streamline coding, enhance automation of extracting Knowledge Graphs, specifically for the use case of user stories.

\section{Research Gap}
Existing researches~\cite{raharjana2021user}\cite{lucassen2017extracting} have demonstrated the feasibility of converting user stories into Knowledge Graphs using Natural Language Processing techniques. However, these approaches often suffer from limitations in precision, and require a high degree of technical expertise to implement. Furthermore, the automation process is not entirely autonomous and still requires significant human intervention, particularly for tasks such as validating extracted components and resolving ambiguities. This highlights the need for a method that achieves a balance between precision and implementation complexity.

A recent study explored the potential of Large Language Models (LLMs) for this task \cite{arulmohan2023extracting}. While promising, this research focused solely on extracting domain knowledge in the form of graph components (nodes and relationships) from user stories and did not implement a solution to generate a Knowledge Graph from the user's perspective- one that would enable graph visualization and querying. Additionally, it relies on a specific Large Language Model, which introduces limitations in terms of model stability, flexibility, and potential bias.

This highlights a significant gap in the understanding of how LLMs can be effectively employed to create a robust and automated solution for extracting Knowledge Graphs from user stories within the broader field of requirements engineering. Addressing this gap can enhance both the accuracy and efficiency of knowledge representation from natural language requirements. Accuracy ensures the extracted Knowledge Graph faithfully represents user stories, while efficiency minimizes the complexity of implementation and computational overhead. This research seeks to balance these aspects, offering a flexible methodology compatible with different language models, ultimately contributing to improved software development practices.

\section{Research Questions}
Understanding user requirements is critical for the development of successful software solutions. This thesis focuses on the potential of employing Large Language Models (LLMs) through the advanced framework LangChain to enhance the process of extracting Knowledge Graph from user stories. Specifically, this research seeks to address the following questions:

\begin{enumerate}
        \item How does the accuracy of nodes and relationships extraction of the proposed solution compare to the existing Large Language Model-based method \cite{arulmohan2023extracting}?

        \item How can a knowledge graph be automatically generated from user stories using large language models without being dependent on a specific provider?

        \item What strategies can be employed to build a fully automated solution to extract knowledge graphs from user stories?
\end{enumerate}

\section{Contributions}
This thesis makes significant contributions to the fields of Requirements Engineering and Large Language Models. The key contributions are summarized as follows:

\begin{enumerate} 
    \item Development of a fully automated approach for extracting knowledge graphs from user stories using \acp{LLM}. This approach simplifies the process of transforming natural language requirements into structured, actionable knowledge, providing a foundation for better understanding and managing software requirements in requirements engineering.
    
    \item Development a reusable evaluation script for assessing the accuracy of the knowledge graph extraction process. This script leverages standard metrics (e.g., precision, recall, F-measure) and relies on a pre-defined ground-truth for objective evaluation.
    
    \item Comparative evaluation of the proposed solution, including its accuracy against existing methods in terms of precision, recall, and F-measure in knowledge graph extraction from user stories.
    
\end{enumerate}

\section{Research Outline}
Chapter~\ref{chap:background} introduces the theoretical concepts that form the foundation of this research. Chapter~\ref{chap:foundation} examines the prior work that this thesis builds upon, serving as a benchmark for results while identifying the limitations that this research aims to address. Chapter~\ref{chap:design} focuses on the design of the \ac{USGT} module and implementation of a fully automated solution for generating knowledge graphs from user stories. Chapter~\ref{chap:eval} outlines the evaluation process, detailing the metrics used, experimental setup, results, and a discussion of the findings. Chapter~\ref{chap:state} reviews the current state of research in this field and highlights potential research gaps. Finally, Chapter~\ref{chap:conclusion} concludes the thesis by summarizing the contributions, addressing the limitations, and proposing directions for future work.

\chapter{Background}
\label{chap:background}
This chapter presents the concepts and theoretical foundation for understanding the extraction of knowledge graphs from user stories using LangChain. This research area sits at the intersection of several key concepts: LLMs, KGs, and user stories. Establishing a solid grasp of these concepts is crucial for effectively framing our research question, methodology, and anticipated contributions.

Section~\ref{sec:re} introduces the field of requirements engineering and highlights its critical role in the software development process. Subsection~\ref{subsec:user_stories} focuses on the significance of user stories within this domain.

Section~\ref{sec:kg} explores knowledge graphs, emphasizing their potential to enhance requirements engineering. Subsection~\ref{subsec:ontology} discusses the ontology reintroduced in this study, followed by Subsection~\ref{subsec:neo4j}, which introduces Neo4j, a graph database used for storing and managing graph structures.

Next, Section~\ref{sec:nlp} delves into \ac{NLP}, presenting common techniques employed to extract structured information from user stories. Section~\ref{sec:llm} highlights \acp{LLM} as a transformative technology in this context.

Section~\ref{sec:langchain} introduces LangChain, detailing its features and how they are leveraged in this study. Lastly, Section~\ref{sec:dataset} describes the dataset utilized in this research for evaluation purposes.

\section{Requirements engineering}
\label{sec:re}
As technology evolves, software development emerges as a critical tool for companies and academia to tackle increasingly complex challenges. From this perspective comes the \ac{RE} that states that well-defined requirements are essential to any system. These requirements capture and communicate characteristics, capabilities, and quality expectations, ensuring the system fulfills its intended purpose and meets user needs \cite{sommerville2011software}. 

A formal definition of a requirement was proposed by the IEEE Standard Glossary of Software Engineering Terminology \cite{ieee1983ieee}:

\textit{\begin{enumerate}
    \item A condition or capability needed by a user to solve a problem or achieve an objective.
    \item A condition or capability that must be met or possessed by a system or a system component to satisfy a contract, standard, specification, or other formally imposed document.
    \item A documented representation of a condition or capability as in 1 or 2.
\end{enumerate}}

The \ac{RE} life cycle \cite{sommerville2011software} is a systematic process that guides the development and management of system requirements from initial conception to final validation. This cycle is essential for ensuring that the system meets all intended goals and user expectations. The \ac{RE} life cycle involves the phases \cite{de2012requirements}:

\begin{itemize}
    \item \textbf{Elicitation}: Identifies stakeholders and captures their needs. This includes collecting business and or user requirements, known as functional requirements, and quality, known as non-functional requirements. 
    
    \item \textbf{Analysis}: Captured requirements are then analyzed in detail during this phase.  This involves decomposing high-level requirements into more specific and actionable ones. Additionally, feasibility is assessed, priorities are negotiated, inconsistencies and conflicts are identified, and unclear, incomplete, ambiguous, or contradictory requirements are resolved.
    
    \item \textbf{Specification}: Following analysis, the well-defined requirements are documented in a clear, consistent, and accessible manner. This specification serves as the baseline for the development process and ensures everyone involved has a shared understanding of the system's functionalities and constraints.
    
    \item \textbf{Verification and validation}: Focuses on ensuring that the documented requirements accurately reflect the needs identified during elicitation and that they meet the project's objectives. Various tests and evaluation methods are employed during verification and validation to confirm that the requirements are complete, consistent, achievable, and traceable throughout the development life cycle.
\end{itemize}

Clear requirements act as a direction for successful software development. They guide each stage of the process, from system design and development to testing, implementation, and ongoing operation.  Misunderstandings at this initial stage can lead to wasted effort and rework later on.  Investing time in thorough planning through well-defined requirements can significantly save time and resources, in the long run, \cite{sommerville2011software}\cite{pohl1996requirements}.

\subsection{User Stories}
\label{subsec:user_stories}
Agile development, a software development paradigm emphasizing flexibility, collaboration, and customer-centricity, emerged from the 2001 Agile Manifesto \cite{beck2001agile}. It promotes iterative development, breaking down projects into manageable units delivered in short cycles. This iterative approach is facilitated by user stories, which capture functionalities from the user's perspective and break them down into manageable development tasks.

User stories are a specific form of user requirements \cite{sommerville2011software}. They are statements that capture desired functionalities, which deliver value to system users or acquirers in a simple natural language. In addition, they can be expressed as a tuple containing the following information \cite{schwaber2011scrum}: 

\begin{itemize}
    \item \textbf{User role}: Also known as persona, it is the user role that has the perspective in the user story.
    \item \textbf{Actions}: They are specific activities that the user role wants to perform within the system.
    \item \textbf{Entities}: They are objects that the actions are performed on or with.
    \item \textbf{Benefit}: It is an optional attribute, that represents the value or advantage that the persona gains by performing the actions over entities.
\end{itemize}

A user story typically follows the Connextra format \cite{cohn2004user}:

\begin{quote}
\textit{As a <\textbf{PERSONA}>, I want <\textbf{ACTIONS} over \textbf{ENTITIES}> so that <\textbf{BENEFIT}>.}
\end{quote}

Scrum \cite{schwaber2011scrum}, a popular Agile framework, heavily relies on user stories. Scrum teams manage a prioritized list of work items called a product backlog. This backlog contains a set of user stories, along with other potential functionalities, bug fixes, or tasks. This emphasis on user stories keeps the focus on delivering value throughout the iterative development process. 

To assess the quality of user stories, the \textbf{INVEST} acronym is commonly employed~\cite{cohn2004user}: 
\begin{itemize}
    \item \textbf{I}: Stands for Independent, meaning that each user story should be self-contained and free of dependencies on other stories, allowing for flexible prioritization and development.
    
    \item \textbf{N}: Represents Negotiable, emphasizing that user stories are not fixed contracts but rather starting points for discussions between stakeholders and the development team.

    \item \textbf{V}: Refers to Valuable, ensuring that each story delivers clear value to the customer or end user.

    \item \textbf{E}: Stands for Estimable, meaning the development team should be able to estimate the effort required to implement the story.

    \item \textbf{S}: Stands for Small, ensuring stories are granular enough to be completed within a single iteration or sprint. 

    \item \textbf{T}: Represents Testable, highlighting the need for clear acceptance criteria that allow for proper validation of the story's implementation
\end{itemize}

By adhering to these principles, user stories can be written in a way that enhances clarity, collaboration, and deliverability within Agile projects.

This thesis research contribution focuses on user stories, thereby contributing to multiple phases of the RE life cycle. In the \ac{RE} life cycle, user stories are fundamentally tied to the specification phase. They effectively capture stakeholders' needs and articulate what users require from the system in a simple and understandable format. However, their utility extends beyond specification. During the elicitation phase, user stories can help to identify stakeholders involved. They also contribute to the analysis phase, where they play a crucial role in prioritizing requirements and identifying dependencies among different system components. Overall, user stories are a versatile and powerful tool in the \ac{RE} process, providing a clear, semi-structured, and concise method for documenting and managing requirements throughout the development life cycle.

\section{Knowledge Graph}
\label{sec:kg}
The concept of a Knowledge Graph is not entirely new. Semantic networks, which are predecessors to \ac{KG}s, emerged in the 1960s as a way to connect pieces of knowledge, laying the foundation for knowledge representation and knowledge-based systems. However, the modern interpretation of \ac{KG}s, characterized by their scale and focus, can be considered a relatively new phenomenon \cite{fenselknowledge}.

As the name suggests, the basis of a \ac{KG} is a graph. A Graph \cite{wicker2002fundamentals}, denoted by G, is a fundamental structure used to model relationships between objects. It consists of two sets: V, the set of vertices, and E, the set of edges that connect these vertices. Also, each edge connects exactly two distinct vertices from V. The next part, Knowledge \cite{fenselknowledge}, is generated by interpreting a graph, and relating its elements to real-world objects and actions.

When both definitions are combined, it is possible to derive that a knowledge graph presents its data and schema in a graph format, in which the vertices (V) represent entities such as people, places, or concepts, and edges (E) represent the relationships between these entities, such as "is located in" or "invented by". 

Knowledge Graphs offer a powerful approach to data management. Their core strength lies in flexible data modeling. Unlike traditional methods, they can easily accommodate data from diverse sources by creating connections between them. This simplifies data integration, making them highly scalable and well-suited for handling information from various sources. 

Particularly for user stories, a \ac{KG} provides an effective solution for modeling by systematically consolidating and organizing information. Diverging from the standard approach that relies on interpreting problem descriptions written in natural language, which can be ambiguous and challenging to manage, \ac{KG}s represent user stories and their attributes as structured data. This structured representation not only clarifies the individual elements of each user story but also highlights the dependencies and relationships between them in a product backlog. By visualizing these connections within a graph framework, \ac{KG}s facilitate a clearer understanding of how user stories interact and impact one another, thereby improving traceability and supporting more efficient management of requirements, and, consequently, of the development process \cite{fenselknowledge}.

\subsection{Ontology}
\label{subsec:ontology}
A \ac{KG} relies on a well-defined underlying ontology to represent its domain \cite{fenselknowledge}. Essentially, an ontology \cite{staab2013handbook} is a formally agreed-upon model of a system's structure that explicitly defines the relevant entities and their relationships.

\begin{figure}[htbp]
  \centering
  \includegraphics[width=0.5\textwidth]{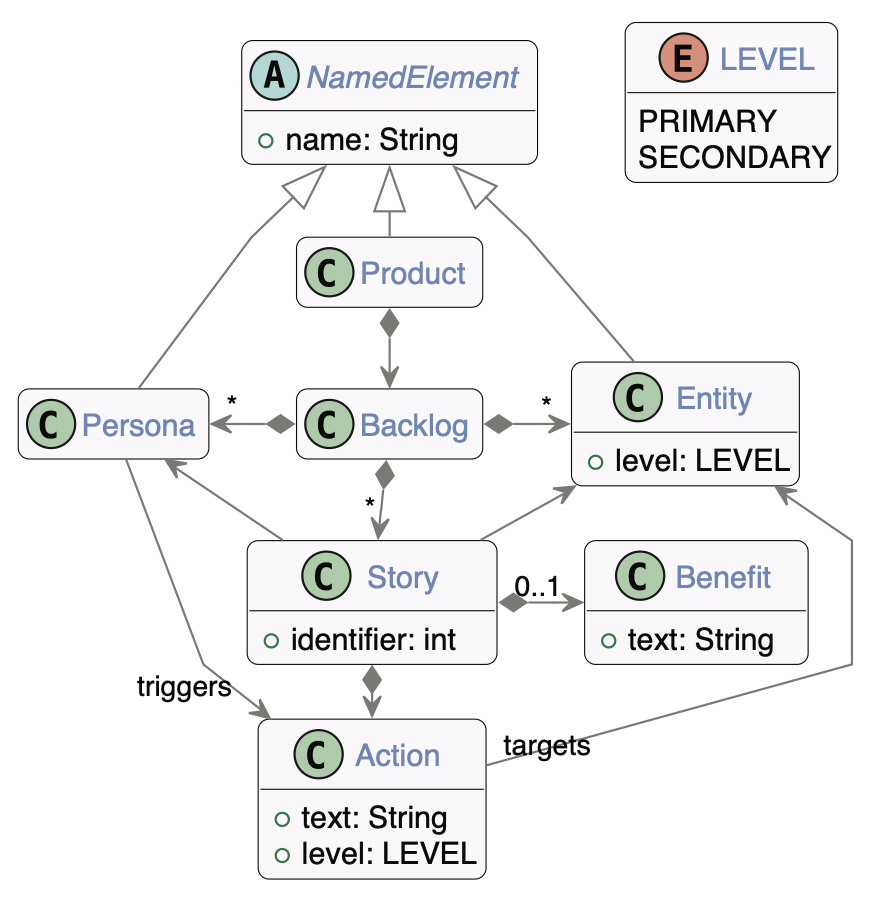}
  \caption{Conceptual model of a backlog \cite{arulmohan2023extracting}.}
  \label{fig:metamodel}
\end{figure}

This thesis proposes a knowledge graph design for representing user stories, also known as backlog items, within the Scrum framework \cite{schwaber2011scrum}. The design adopts the ontology developed by Arulmohan et al.~\cite{arulmohan2023extracting} (see Figure~\ref{fig:metamodel}), provides a structured approach to define user stories within a KG. 

In the metamodel, each product consists of one or more backlogs, where each backlog corresponds to a set of \textit{N} user stories. Each user story is associated with a persona, and at least one entity, and is composed of at least one action. Additionally, a user story can optionally include one benefit. A persona is associated with the primary level action (also understood as the main action) through a \textit{triggers} relationship, and each action is associated with an entity through a \textit{targets} relationship. From the backlog perspective, it encompasses a set of user stories along with the personas, entities, actions, and possibly benefit associated with those stories.

To illustrate, consider the scenario in which the product is a web application. The user stories are organized in one backlog referring to the system administration capabilities of this web application, and they reflect the following user requirements:  

\begin{quote}
\textit{US1. As a user, I want to access my personal data, so that I can review it.}
\end{quote}

\begin{quote}
\textit{US2. As an administrator, I want to be able to see the user's email, so that I can assign a new role.}
\end{quote}

\begin{quote}
\textit{US3. As a user, I want to see my own role.}
\end{quote}

\begin{figure}[ht]
  \centering
  \includegraphics[width=0.9\textwidth]{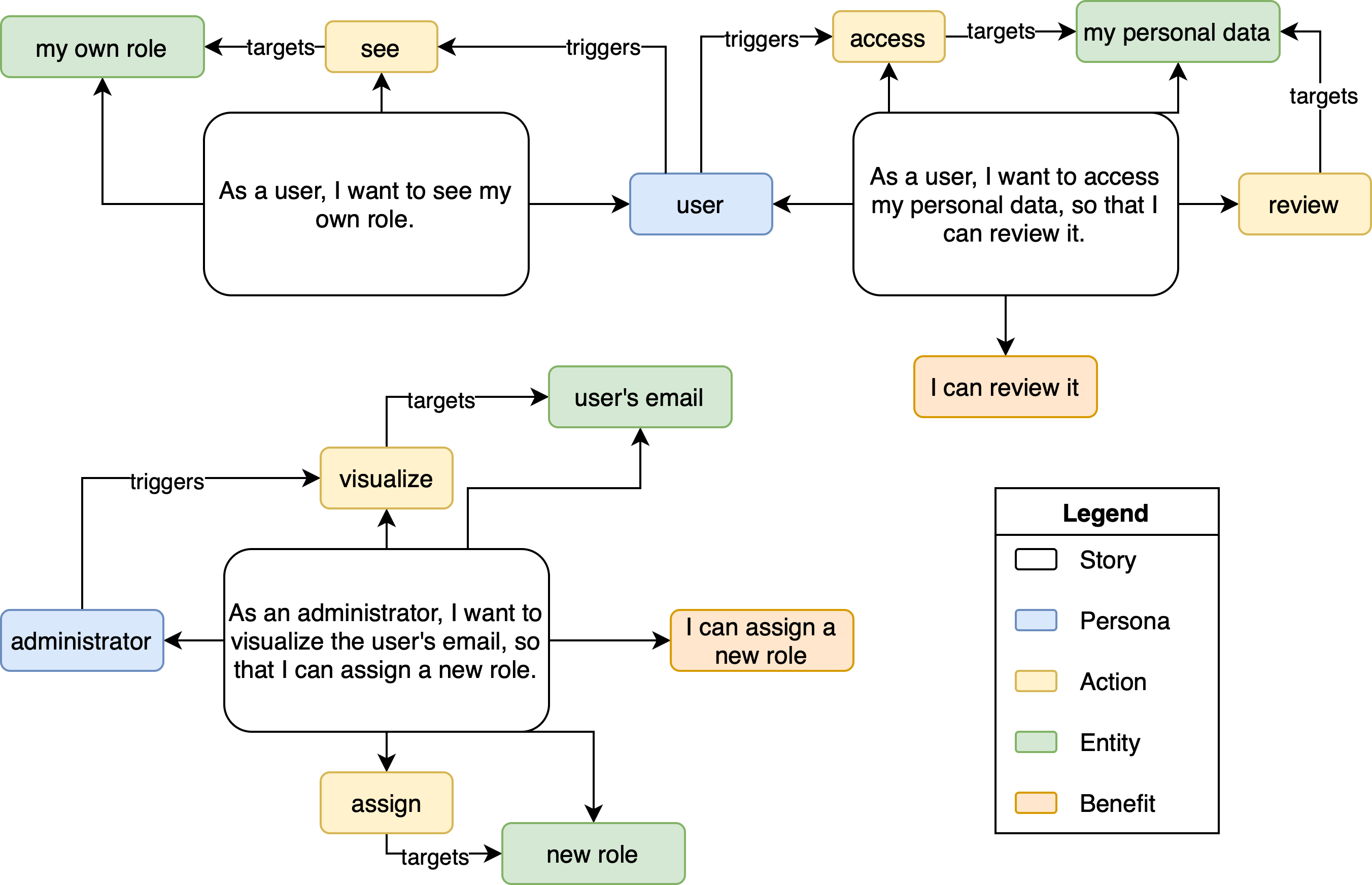}
  \caption{Example of a modeled product backlog of a web application following the ontology described in Figure~\ref{fig:metamodel}.}
  \label{fig:ontology_applied}
\end{figure}

These user stories can be modeled according to the ontology described on the metamodel in Figure~\ref{fig:metamodel}, resulting in the knowledge graph seen in Figure~\ref{fig:ontology_applied}. The nodes are classified into 5 types: \textit{story}, \textit{persona}, \textit{action}, \textit{entity}, and \textit{benefit}. The relationships from the user story to the remaining node types guarantee traceability, while the \textit{triggers}, and \textit{targets} relationships provide insights into the focus of requirements and can help prioritize their development.

In the example provided, it becomes easier to identify relationships and dependencies that may not be immediately obvious in textual form, such as the connection between the \textit{persona} in US1 and US3, which can help prioritize development tasks, improve role-specific features, and ensure consistency across stories. Changes to the functionalities associated with a particular persona can be analyzed more effectively. For instance, if a feature related to the \textit{user} persona needs modification, the graph immediately highlights which user stories and corresponding actions might be impacted. Furthermore, the graph improves communication between stakeholders, as it becomes easier to track which personas interact with specific features and identify potential gaps or overlaps in the backlog.

Even with a simple example, it is already possible to see that adopting this ontology can ensure a consistent and well-defined structure for representing user stories within \acp{KG}. This structured representation facilitates reasoning, analysis, and automation of tasks related to user stories.

\subsection{Neo4j Database}
\label{subsec:neo4j}
Neo4j~\cite{neo4j} is a popular graph database product specifically designed to store and manage interconnected data. Opposed to some relational databases that try to mimic graph structures, Neo4j is built from the ground up to handle connected data efficiently. This native approach allows for faster performance and more natural graph queries.

Neo4j utilizes Cypher, a declarative query language specifically designed for graph data. Cypher allows you to write intuitive queries that reflect the relationships within your data, simplifying analysis compared to using SQL in relational databases.

In this thesis, Neo4j was selected as the tool for storing and visualizing the resulting \ac{KG} due to its robust capabilities in modeling entities and relationships. Neo4j excels in representing complex connections within data, making it ideal for capturing the structure of user stories and their dependencies. Additionally, its architecture supports easy scalability, which is important considering the chosen dataset. Neo4j also seamlessly integrates with the existing technical setup, ensuring smooth implementation and operation within the project's framework.

\section{Natural Language Processing}
\label{sec:nlp}
\textit{\acf{NLP}}, a subfield of \ac{AI}, focuses on facilitating the interaction between computers and human language \cite{nadkarni2011natural}. Among its primary tasks is \textit{\ac{IE}}, which involves the automatic conversion of unstructured text into structured data \cite{cambria2014jumping}. This structured information can then be leveraged for diverse applications, including populating knowledge graphs.

\ac{NLP} techniques present considerable advantages for enhancing the quality of user stories. Through the application of \ac{NLP}, it is possible to effectively parse, extract, and analyze data from user stories. \ac{NLP} has seen widespread use in the software engineering domain, addressing a variety of tasks. These techniques include \cite{nadkarni2011natural}:

\begin{itemize}
    \item \textbf{Tokenization}: This fundamental step breaks down a text into individual units, typically words or sentences.
    
    \item \textbf{\ac{POS} tagging}: This process assigns a grammatical category (e.g., noun, verb, adjective) to each word (or tokens) in a text.
    
    \item \textbf{Dependency Parsing}: This technique goes beyond \ac{POS} tagging by analyzing the grammatical relationships between words in a sentence. It identifies how words depend on each other syntactically.
    
    \item \textbf{\ac{NER}}: This subtask involves identifying and categorizing named entities in a text into predefined categories (e.g. names of people, organizations, or locations). 
\end{itemize}

These \ac{NLP} techniques can be applied to user stories to extract key information, as well as help improve user story clarity by identifying ambiguous language or inconsistencies. For instance, \ac{POS} tagging can help identify user roles (nouns) and actions (verbs), while \ac{NER} can distinguish specific functionalities mentioned in the story.  This focus on clarity ensures a better understanding between stakeholders and development teams.

These diverse applications highlight the potential of \ac{NLP} to transform user stories from unstructured to structured information. Recent advancements~\cite{raharjana2021user} show particular promise in automating the modeling of information and extracting key abstractions.

\section{Large Language Models}
\label{sec:llm}
Within the field of \ac{NLP} and \ac{AI}, \acp{LLM} stand out as powerful tools. These models are essentially statistical methods trained on massive amounts of text data.  Through this training, \acp{LLM} learn the underlying statistical patterns and relationships between words within a language.  This capability allows them to predict the likelihood of specific word sequences, enabling a wide range of \ac{NLP} applications \cite{mcteartransforming}.

\acp{LLM}~\cite{chang2023survey} are a type of advanced neural network architecture specifically designed for processing and generating human language. They are characterized by huge amounts of parameters, which allows them to capture complex relationships and patterns within language. In addition, they have the ability to learn and adapt to various language tasks.

\begin{table}[htbp]
  \centering
  \begin{tabularx}{\textwidth}{|p{2.5cm}|X|X|}
    \hline
    \textbf{Concept} & \textbf{Description} \\
    \hline 
    Temperature &  Temperature controls the stochasticity of the model's output. It varies from zero to one, lower values favor statistically frequent and predictable responses.\\
    \hline
    Role & The role describes who is interacting with an LLM. The standard roles are "user" (gives input), "assistant" (gives output), "system" (sets context), and "tool" (access external APIs or tools). \\
    \hline
    Hallucination & It is the phenomenon when generated responses are factually incorrect and do not correspond to real-life information. \\
    \hline
    Tokens & Tokens are the smallest units of information in the form of text the LLM can handle. It is an enumeration and can be a word, subword, and even punctuation.\\
    \hline
    Prompt &  A prompt is a set of instructions or an initial piece of information that guides the LLM toward a desired outcome.\\
    \hline
    Prompt engineering & It is the process of designing and improving prompts to achieve specific outputs. \\
    \hline
    Instruction tuning & It is the process of fine-tuning the LLM by training it on a dataset of instructions to achieve desired outputs.\\
    \hline
    In-context learning & Enables the LLM to learn new skills by incorporating examples of the desired output within the prompt.\\
    \hline
    Zero-Shot prompt & It is a technique based on Zero-Shot Learning, that uses natural language text to obtain a desired output without providing examples. \\
    \hline
    Few-Shot prompt & It builds on the zero-shot prompting, but it actually provides examples of the expected output. \\
    \hline
  \end{tabularx}
  \caption{Fundamental terms for understanding LLMs.}
  \label{tab:llm_terms}
\end{table}

Table~\ref{tab:llm_terms} presents essential terminology to comprehend \acp{LLM}, as adapted from~\cite{fan2023large}. While concepts like \textit{role} are specific to LLM-based chat applications, other terms such as temperature, prompt, and prompt engineering are broadly applicable to a wide range of LLM applications, including but not limited to chat-based systems.

The fundamental component of \acp{LLM} is the transformer architecture \cite{chang2023survey}. This architecture was designed to address the limitations of previous sequence-to-sequence models in NLP, which typically relied on \acp{RNN} and \ac{LSTM} networks. These earlier models processed data sequentially and struggled with long-range dependencies and parallelization. In contrast, the transformer uses a novel mechanism called self-attention to handle sequences, enabling more efficient processing and better performance on a wide range of tasks \cite{vaswani2017attention}.

A notable capability of \acp{LLM} is in-context learning \cite{mcteartransforming}. This refers to the model's ability to enhance its understanding and response to prompts by incorporating additional contextual information directly into the prompt itself. This technique allows \acp{LLM} to adapt to new tasks with minimal additional training, simply by providing relevant examples or supplementary data within the prompt \cite{chang2023survey}. This makes \acp{LLM} particularly versatile and useful for applications where the model needs to adapt quickly to new and varied tasks.

Prompt engineering is a common used technique for interacting effectively with \acp{LLM}, particularly in applications like chatbots and virtual assistants \cite{chang2023survey}\cite{unleashingGPT}.This approach involves strategically crafting input prompts that guide the \ac{LLM} towards generating accurate and relevant responses. By effectively translating user intent into a format that \acp{LLM} can process, prompt engineering optimizes interactions and significantly enhances the relevance and quality of the model's outputs.

In the context of \ac{RE}, prompts can be used as natural language instructions given to an \ac{LLM} to perform specific tasks, such as identifying whether a certain system functionality is properly elicited from user stories \cite{white2024chatgpt}, which can improve efficiency of tasks and reduce human effort.

Large Language Models have demonstrated exceptional proficiency in processing and understanding natural language, making them invaluable tools across a wide range of tasks. Their usability, even for non-technical users, underscores their potential as a groundbreaking technology suitable for diverse applications. This thesis aims to delve into the capabilities of \acp{LLM}, specifically focusing on their ability to automatically generate knowledge graphs from user stories. By utilizing the natural language understanding and generation strengths of \acp{LLM}, this research evaluates how effectively these models can extract and organize requirements from user stories into structured knowledge representations. 

\section{Langchain}
\label{sec:langchain}
While \acp{LLM} hold significant potential for various applications, their implementation in real-world scenarios can be complex and challenging, due to factors such as choosing the right language model, designing effective prompts, and managing data processing workflows. LangChain~\cite{langchain-git}, an open-source Python framework, addresses these complexities by simplifying the development of LLM-powered applications, by providing clear connectivity to various third-party LLMs (e.g., OpenAI, Google Gemini, Ollama), and by offering a declarative syntax called \ac{LCEL} to manage complex \ac{LLM} interactions, and therefore becoming a strong candidate to address the limitations of previous researches.

\ac{LCEL} \cite{langchain-git} is the declarative syntax provided by LangChain that simplifies the implementation for users and handles the underlying execution details. It offers several key benefits: first, \ac{LCEL} allows for rapid prototyping with facilitated transition to production due to its flexibility and lack of code changes needed. Second, it optimizes data processing through features like streaming support, asynchronous execution, and parallel processing, leading to faster and more efficient workflows. Finally, \ac{LCEL} enhances chain reliability with retries, fallbacks, and access to intermediate results, while also ensuring data integrity through automatic input and output schema generation. These combined features make \ac{LCEL} a powerful tool for building robust and scalable \ac{LLM} applications within the LangChain framework.

The idea behind this framework is to build custom chains, and this can be achieved by the Runnable protocol \cite{langchain-git}. Designed to execute a specific task or series of operations. It acts as a modular component that can be composed with others to build custom workflows, where each Runnable performs a defined action, such as processing input data or interacting with external systems. One key operation that a Runnable supports is the invoke method, which allows it to process a single input and return the corresponding result. This method is central to executing tasks within the chain, enabling the flow of data and the execution of operations in a sequential manner. The components in Table~\ref{tab:langchain_components} are examples of Runnables.

\begin{table}[htbp]
    \centering
    \begin{tabularx}{\textwidth}{|p{2.5cm}|X|X|X|}
        \hline
        \textbf{Component} & \textbf{Description} & \textbf{Input Type} & \textbf{Output Type} \\
        \hline
        Prompt Template & Transform user inputs and parameters into instructions for a language model & Dictionary & Prompt Value \\
        \hline
        ChatModel & Language Models that support distinct roles in conversations (e.g. AI, human, system) & Single string, list of chat messages or a Prompt Value & ChatMessage \\
        \hline
        OutputParser & Process the output of a ChatModel into a desired format & The output of a ChatModel & Depends on the parser (e.g., String, JSON) \\
        \hline
    \end{tabularx}
    \caption{LangChain Components.}
    \label{tab:langchain_components}
\end{table}

LangChain itself doesn't host any language models but relies on integration with third-party options. Additionally, the framework offers standardized parameters applicable to any LLM, such as model name, temperature, and maximum tokens generated.

Prompt templates are another key feature, guiding a model's response. They take a dictionary as input, with each key representing a variable in the prompt template to fill in. By providing a structured format for inputs, prompt templates help standardize and refine interactions with these models.

An example in Listing~~\ref{lst:langchain_example} demonstrates how LangChain simplifies the creation of standardized interactions with \ac{LLM}s using minimal code. This example chain consists of three key components: a ChatModel to generate responses, a Prompt template to standardize the input format, and a String Output Parser to process the model's output, removing all the metadata that comes with the output, and keeping only the model's reply. The chain relies on the pipe operator, a core feature of the \ac{LCEL} syntax, to connect components sequentially and ensure the data flow remains clear and readable.

\begin{lstlisting}[language=Python, caption={Example of LangChain application.}, label={lst:langchain_example}, float]
from langchain_openai import ChatOpenAI
from langchain_core.output_parsers import StrOutputParser
from langchain_core.prompts import ChatPromptTemplate

model = ChatOpenAI(model="gpt-3.5-turbo-0125")
prompt = ChatPromptTemplate.from_template("what is the capital of {country}?")
chain = prompt | model | StrOutputParser()

reply = chain.invoke({"country": "Brazil"})
\end{lstlisting}

LangChain enhances the capabilities of language models by consistently integrating them with external data sources and computational systems. This connectivity allows language models to act as reasoning engines while delegating tasks such as knowledge retrieval and execution to other specialized systems. A prime example of this is LangChain's integration module for the Neo4j graph database. This module facilitates querying, updating, and even constructing knowledge graphs, aligning perfectly with the objectives of this research.

In this thesis, LangChain was chosen as the development framework due to its features that directly align with this research objective of exploring LLM-based knowledge graph generation from user stories. A key advantage of LangChain is its ability to fully automate the creation of knowledge graphs from user stories, from extracting nodes and relationships to processing the LLM's output and storing it in a graph database. This practical aspect significantly improves the usability of the approach, especially for less technical users who may not have extensive experience with the technologies involved in modeling user stories. These users can rely on the assistance of \ac{LLM}s to simplify this task through LangChain.  Notably, LangChain offers pre-built modules for knowledge graph creation and facilitates seamless integration with graph databases, particularly Neo4J. This integration is crucial for storing and analyzing the generated knowledge graphs.

\subsection{LLMGraphTransformer}
\label{subsec:llm_transformer}
Converting unstructured text data into structured knowledge graphs offers a powerful approach to gaining deeper insights and navigating complex relationships. The \textit{LLMGraphTransformer} module \cite{langchain-git} within LangChain plays a crucial role in this process. It utilizes \acp{LLM} to analyze text documents and extract key entities (nodes) and the connections between them (relationships). These extracted elements are then organized into a structured graph format, facilitating efficient analysis and exploration of the information. It is important to remark that this module is part of the LangChain Experimental code, and, therefore, it may lack extensive testing, and is designed for research and experimental use only.

To achieve its goal, it uses a predefined prompt specifically designed to guide the LLM toward extracting relevant entities and relationships suitable for constructing general-purpose Knowledge Graphs (KG). In addition, it has an end-to-end data processing logic that handles the entire data process pipeline, transforming raw text into a KG document.

The input of this module is a Document~\cite{langchain-git}, which serves as the fundamental unit of information. It comprises two key components: page content and metadata. The page content is the core textual content that interacts with the language model, in the context of this thesis, the page content specifically represents the user story input. The metadata is an optional field, which provides additional information about the unit of information.

The module interacts with the LLM using its predefined prompt through a chain and
receives a response in string format. Then, it employs techniques to convert the Nodes and Relationships received in raw string format into objects of classes Node and Relationship respectively, and lastly, the KG document is created. The end-to-end processing capability simplifies the knowledge graph construction process by eliminating the need for manual intervention in data transformation steps.

However, it is important to note that the effectiveness of LLMGraphTransformer depends heavily on the chosen \ac{LLM}.  Since \ac{LLM}s are statistical models, the accuracy and nuance of the extracted graph data can vary depending on their specific capabilities.

While LLMGraphTransformer is designed to convert text into knowledge graphs, it often prioritizes high-level entities like names, places, and companies during extraction. This focus can lead to missing crucial details and entities specific to user stories, such as actions, processes, or domain-specific concepts. These entities, even if not proper names or locations, are vital for a comprehensive user story \ac{KG}, and therefore this limitation will be addressed in the proposed solution design and implementation.

\subsection{Neo4j integration}
\label{subsec:neo4j_int}

\begin{lstlisting}[language=Python, caption={Neo4j and LangChain integration}, label={lst:neo4j_example}, float]
from langchain_community.graphs import Neo4jGraph
import os

os.getenv("NEO4J_URI")
os.getenv("NEO4J_USER")
os.getenv("NEO4J_PASSWORD")

graph = Neo4jGraph()
\end{lstlisting}

The final step in building a knowledge graph is integrating it with a graph database for storage, with Neo4j being the chosen database in this study. LangChain simplifies this process through its Neo4jGraph module, which provides seamless integration with Neo4j.

Listing~\ref{lst:neo4j_example} illustrates a basic example of connecting to a Neo4j instance, using environment variables for secure authentication. Through the Neo4jGraph module, users can easily create a graph instance and populate a knowledge graph within Neo4j, by providing a specific data type, the Graph Document. This module significantly streamlines the interaction between the LangChain framework and the graph database, reducing complexity and enhancing automation.

The integration with Neo4j allows for not only the storage but also the further analysis of the generated knowledge graph. By using a graph database, users can run queries and perform analysis on the knowledge graph, enhancing its value and utility in the requirements engineering process. This is a core advantage of the proposed approach in comparison to previous researches.

\section{Dataset Description}
\label{sec:dataset}
The dataset used in this research, named Ace-design/qualified-user-stories-dataset, was curated by Sathurshan Arulmohan, Sébastien Mosser, and Marie-Jean Meurs \cite{qualified-user-stories_2023}. It builds upon the original dataset published by Dalpiaz et al. \cite{dalpiaz_requirements_2018}, which consists of 22 product backlogs containing 1679 user stories. The Ace-design dataset is an annotated version of the original, enriched with additional metadata and annotations for enhanced analysis and modeling purposes.

\begin{table}[htbp]
  \centering
  \begin{tabular}{|c|c|}
    \hline
    \textbf{Backlog file name} & \textbf{Context} \\
    \hline
    g02   & Reporting \\
    \hline
    g03   & Management application \\
    \hline
    g04   & Management application \\
    \hline
    g05   & Reporting \\
    \hline
    g08   & Development \\
    \hline
    g10   & Web \\
    \hline
    g11   & Web \\
    \hline
    g12   & Management application \\
    \hline
    g13   & Web \\
    \hline
    g14   & Content management \\
    \hline
    g16   & University \\
    \hline
    g17   & Development \\
    \hline
    g18   & Content management \\
    \hline
    g19   & Internet of Things \\
    \hline
    g21   & Management application \\
    \hline
    g22   & Content management \\
    \hline
    g23   & Content management \\
    \hline
    g24   & Content management \\
    \hline
    g25   & Content management \\
    \hline
    g26   & Content management \\
    \hline
    g27   & Content management \\
    \hline
    g28   & Management application \\
    \hline
  \end{tabular}
  \caption{Overview of dataset content.}
  \label{tab:backolog_file}
\end{table}

The user stories were meticulously annotated by a systematic process that involved using Doccano \cite{doccano}, an open-source text annotation tool, to recognize and label relevant entities (i.e., Persona, Action, Entity, Benefit). This process was followed by rigorous quality assurance measures, including calibration on a subset of stories, regular meetings to discuss annotations, and manual review of randomly selected stories \cite{arulmohan2023extracting}.

Each backlog is structured in a distinct JSON file format, which file name and context can be seen in Table~\ref{tab:backolog_file}, where each file contains a list of dictionaries. Each dictionary represents a user story along with its annotated components, as seen in Listing~\ref{lst:json_example}.

\begin{lstlisting}[caption={Example of annotated user story in JSON format from g02 backlog.}, label={lst:json_example}, float]
{
    "PID": "#G02#",
    "Text": "#G02# As a Website user, I want to access published FABS files, so that I can see the new files as they come in.",
    "Persona": [
        "Website user"
    ],
    "Action": {
        "Primary Action": [
            "access"
        ],
        "Secondary Action": [
            "see"
        ]
    },
    "Entity": {
        "Primary Entity": [
            "published FABS files"
        ],
        "Secondary Entity": [
            "new files"
        ]
    },
    "Benefit": "I can see the new files as they come in",
    "Triggers": [
        [
            "Website user",
            "access"
        ]
    ],
    "Targets": [
        [
            "access",
            "published FABS files"
        ],
        [
            "see",
            "new files"
        ]
    ],
    "Contains": []
}
\end{lstlisting}

As a result of this annotation process, the level of consistency between the different evaluators achieved an impressive 94~\% \cite{arulmohan2023extracting},  which significantly endorsed the selection of this dataset for use in this thesis. This high level of reliability assured the quality and consistency of the annotated user stories, validating their suitability for the research objectives and ensuring robust and dependable results.

\chapter{Prior Research as a Comparative Foundation}
\label{chap:foundation}
This section presents an overview of the research work by Sathursan Arulmohan, Marie-Jean Meurs, and Sébastien Mosser, titled \textit{Extracting Domain Models from Textual Requirements in the Era of Large Language Models}~\cite{arulmohan2023extracting} that forms the foundation of the approach proposed in this thesis. Besides the outstanding contribution of their research in annotating a comprehensive user stories dataset~\cite{qualified-user-stories_2023}, their work is pioneering in employing a distinct technique from traditional \ac{NLP} methods to model user stories by applying \acp{LLM}, specifically Chat-GPT \cite{openai_chatgpt} by OpenAI. 

\section{Research Goal}
This research addresses a key challenge in the context intersection between \ac{RE} and Agile methodologies, which is to extract domain models from user stories. The idea is that by extracting domain models, the ambiguity and conflicts of the requirements can be minimized.

When performed manually, this process has a high chance of errors and takes a long time to complete. To address these challenges, the authors explore the potential of \acp{LLM} and compare their capabilities to both a state-of-practice requirements extraction tool and a dedicated \ac{NLP} approach.

This research represents an initial step toward the use of LLMs to automate requirements engineering tasks. By examining the performance of these tools on a real-world dataset, the paper lays the groundwork for future advancements in integrating artificial intelligence into the RE process, particularly in Agile environments where textual requirements dominate.

\section{Background}
This solution proposes a specific ontology to structure the domain model (which was previously presented in Figure~\ref{fig:metamodel}). It applies the concepts of agile methodology \cite{beck2001agile} that elicits requirements as a set of prioritized user stories, also known as backlog to structure this ontology.

To standardize and make the API's output useful for further processing and evaluation, the solution uses \textit{function calls}, a feature introduced by OpenAI that extends the capabilities of their language models by controlling the model's output. Instead of asking the system to generate free-form text, developers can provide a specific structure for the output. This structure, defined using a JSON schema, acts as a template and can be seen in Listing~\ref{lst:function_call_ex}. The system then fills in the template with appropriate data, ensuring the response matches the exact format needed.

\begin{lstlisting}[caption={\textit{function call} Example \cite{arulmohan2023extracting}.}, label={lst:function_call_ex}, float]
record_extracted_concepts = { 
        "name": "record_elements",
        "description": "Record the elements extracted from a story",
        "parameters": {
            "type": "object",
            "properties": {
                "personas": {
                    "type": "array",
                    "description": "The list of personas extracted from the story",
                    "items": { "type": "string" }
                },
                "entities": {
                    "type": "array",
                    "description": "The list of entities extracted from the story",
                    "items": { "type": "string" }
                },
                "actions": {
                    "type": "array",
                    "description": "The list of actions extracted from the story",
                    "items": { "type": "string" }
                },
                "benefit": {
                    "type": "string",
                    "description": "A single string containing the benefit expected from the story",
                },
            },
            "required": ["personas", "entities", "actions", "benefit"],
        }
      }
\end{lstlisting}

When the API uses a \textit{function call} to produce an output, the interaction is immediately terminated, not allowing the sequence of prompts necessary to complete the three necessary steps to extract user stories information: extract concepts, categorize them, and extract relations. To overcome this limitation, the approach engaged in a conversation with the model to collect each answer in a dedicated schema, an example can be seen in Listing~\ref{lst:api_call_ex}.

\begin{lstlisting}[caption={Prompt Definition and API Call Example \cite{arulmohan2023extracting}.}, label={lst:api_call_ex}, float]
import openai

# prompt definition
conversation = {'role': 'system', 'content':'The elements you are asked to extract from the stories are the following: Persona, Action, Entity, Benefit. A Story can contain multiple elements in each category.'}

# API call
response = openai.ChatCompletion.create(
      model       = model,
      functions   = [record_extracted_concepts],
      messages    = conversation,
      temperature = 0.0 
   )
\end{lstlisting}

\section{Prompt Engineering and API Usage}
The solution proposed is based on prompt engineering to induce the \ac{LLM} to produce the desired result, in this case, to extract concepts and relationships within a user story. 

Four different prompts were created to tackle the distinct steps: system setup, extract concepts, categorize them, and extract relations:

\begin{enumerate}
    \item \textbf{System setup}: the system role is impersonated, instructing the language model to take on a specific character. The overview of the task is presented.

    \item \textbf{Extract concepts}: continuing in the system role, the task is refined to involve detailed extraction of personas, entities, actions, and benefits from a user story. A concrete example of this extraction process is provided, referencing a previously annotated story. Subsequently, the user role is assumed, presenting the user story to be processed.

    \item \textbf{Categorize concepts}: given the model's inability to retain information across interactions, the previously derived concepts must be explicitly reintroduced into the dialogue. Adopting the assistant role, a conversational entry is added, detailing the extracted concepts. Subsequently, the system role is impersonated to specify the task of classifying primary and secondary actions and entities, accompanied by an example.

    \item \textbf{Extract relations}: in the last step, the assistant role introduces the categories. Then, the system role task is described and an example is given.
\end{enumerate}

Then, user stories were fed one by one to the API to avoid hallucinations and overcome the tokens limitation in OpenAI's API.

\section{Experimental Setup}
In their experiment, they compare their F-measure (harmonic mean of precision and recall) results with two other approaches: Visual Narrator, and Conditional Random Fields. The F-measure metric is chosen because it accounts for both correctly and incorrectly classified observations. An F-measure score of 1.0 indicates a perfect match, while a score of 0.0 signifies that either precision or recall is completely lacking.

Visual Narrator \cite{robeer2016automated} is open-source software that uses NLP-based techniques, such as \ac{POS} tagging and rule-based extraction to extract conceptual models from user stories.

\ac{CRF} \cite{lafferty2001conditional} is a \textit{Markov Random Fields} class, a statistical model that can be used to predict patterns based on their context. To apply \ac{CRF} for this specific task, 20\% of the dataset was used to train the model, by applying \ac{POS} tagging to each word in a user story to label the data. Once trained, the model could predict the semantic role of unseen user stories.

To ensure a fair comparison across the three tools, the annotated dataset \cite{qualified-user-stories_2023} was cleaned to include only the user stories that all tools could process. This resulted in a subset of 1,459 user stories, representing 87\% of the complete dataset. In addition, since Visual Narrator is not able to identify relations, the trigger and target relations were not evaluated.

To accurately compare results, a criterion was determined, resulting in the following three comparison modes: 

\begin{itemize}
    \item \textbf{Strict}: In the strictest sense, a result is considered perfect when the AI produces elements that exactly match those in the ground truth. An example of this comparison mode can be seen in Table~\ref{tab:strict_comparison}.

    \item \textbf{Inclusive}: To introduce some flexibility, a scenario where the AI output is a superset of the ground truth was considered. Here, we check that the baseline elements are included within the AI-generated result. This approach allows for a broader range of correct outputs. An example of this comparison mode can be seen in Table~\ref{tab:included_comparison}.

    \item \textbf{Relaxed}: Further relaxing the constraints, the Relaxed comparison mode allows for equivalencies such as treating plurals and singulars as the same or ignoring adjective qualifiers. This approach accounts for variations in how concepts might be expressed. An example of this comparison mode can be seen in Table~\ref{tab:relax_comparison}.
\end{itemize}

\begin{table}[htbp]
  \centering
  \begin{tabular}{|c|c|c|}
    \hline
    \textbf{Ground truth} & \textbf{Experiment results} & \textbf{Comparison results} \\
    \hline
    webpage & webpages & False\\
    \hline
    all webpages & webpages & False\\
    \hline
    user's webpage & [user, webpage] & False\\
    \hline
    webpage & webpage & True\\
    \hline
  \end{tabular}
   \caption{Strict comparison example.}
    \label{tab:strict_comparison}
\end{table}

\begin{table}[htbp]
  \centering
  \begin{tabular}{|c|c|c|}
    \hline
    \textbf{Ground truth} & \textbf{Experiment results} & \textbf{Comparison results} \\
    \hline
    webpage & webpages & True\\
    \hline
    all webpages & webpages & False\\
    \hline
    user's webpage & [user, webpage] & False\\
    \hline
    webpage & webpage & True\\
    \hline
  \end{tabular}
   \caption{Inclusive comparison example.}
    \label{tab:included_comparison}
\end{table}

\begin{table}[htbp]
  \centering
  \begin{tabular}{|c|c|c|}
    \hline
    \textbf{Ground truth} & \textbf{Experiment results} & \textbf{Comparison results} \\
    \hline
    webpage & webpages & False\\
    \hline
    all webpages & webpages & True\\
    \hline
    user's webpage & [user, webpage] & False\\
    \hline
    webpage & webpage & True\\
    \hline
  \end{tabular}
   \caption{Relaxed comparison example.}
    \label{tab:relax_comparison}
\end{table}

The results showed that while Visual Narrator performed reasonably well in identifying personas and actions due to their predictable positions within the text, it struggled with entity extraction, often misidentifying or omitting them.

The proposed solution demonstrated a substantial improvement over Visual Narrator in all categories. It required significantly less development effort and achieved superior results, particularly in identifying actions.

However, despite their overall strength, their solution was outperformed by a CRF-based model, which was trained specifically for this task. Still, the ChatGPT-based solution was considered a promising approach for rapid prototyping.

\section{Discussion}
The work of Arulmohan et al. serves as a crucial foundation for this research due to its pioneering application of \acp{LLM} to the domain of user story modeling. By demonstrating the feasibility of extracting domain models from textual requirements using \acp{LLM}, their research provides a strong starting point for our exploration. 

While their work lays the essential groundwork, this thesis aims to extend their findings by addressing the limitation of their approach, which is heavily reliant on a specific LLM provider (OpenAI) and therefore raises concerns about model stability, potential biases, and the impact of model updates on the overall solution. By exploring a model-agnostic approach, this thesis intends to enhance the scalability and robustness of user story modeling.

Another limitation of their work lies in its exclusive focus on concept extraction from user stories. While their experiment results in components of a \ac{KG} in a JSON format, it does not provide a graph representation from a practical user's perspective. This means that additional implementation steps are required to enable graphical visualization and querying of the information. By neglecting the implementation of knowledge graphs which would allow users to visualize, query, and analyze the data, the research fails to capitalize on the full potential of user stories for enhancing the software development process. This thesis addresses this limitation by introducing a methodology for automatically generating and implementing knowledge graphs from user stories, thereby providing a more comprehensive understanding of system requirements.

\chapter{Design and Implementation}
\label{chap:design}
The previous chapters established the motivation, theoretical foundation, and basis for this research by exploring user stories, knowledge graphs, and large language models. Building upon this groundwork, this chapter delves into the practical aspects of the research by detailing the design and implementation and paves the way for the evaluation and analysis of its accuracy in the following chapter. 

This chapter begins in Section~\ref{sec:preliminary} by describing a preliminary experiment using LangChain's default module \textit{LLMGraphTransformer} to extract a \ac{KG} from user stories, highlighting its limitations and the need for a customized approach. Section~\ref{sec:usgt_design} introduces the tailored module, the \ac{USGT}, explaining how its architecture aligns with the defined ontology (Subsection~\ref{subsec:kg_architecture}) and the role of prompt design in its functionality (Subsection~\ref{subsec:prompt}). The subsequent Section~\ref{sec:automated_design} outlines the design of the complete solution, from processing a product backlog to utilizing the \ac{USGT} module and storing the resulting knowledge graph in a Neo4j database. Lastly, Section~\ref{sec:usgt_implementation} covers the implementation details of the \ac{USGT} module.

\section{Preliminary Experiment}
\label{sec:preliminary}
In the initial phase, an experiment was set up using the default \textit{LLMGraphTransformer} module~\cite{llmgraph} to explore the feasibility of using this module as a solution for extracting a Graph Document from user stories. Even though this module is quite versatile, it was designed for general-purpose knowledge graphs, and therefore it struggled to fully capture user stories nuances.

\begin{figure}[htbp]
  \centering
  \includegraphics[width=0.5\textwidth]{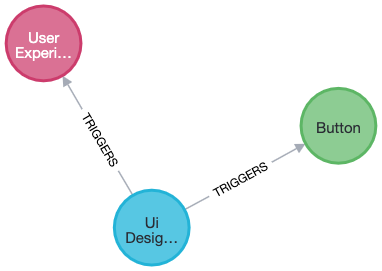}
  \caption{Neo4j visualization of the graph created in Listing~\ref{lst:default_llm}.}
  \label{fig:neo4j_default_llm}
\end{figure}

To exemplify the difficulty of the default \textit{LLMGraphTransformer}, an example is given in Listing~\ref{lst:default_llm} and the results can be seen in Figure~\ref{fig:neo4j_default_llm}. The \ac{LLM} was able to identify only one node correctly (\textit{UI designer} as Persona) and two nodes partially correctly (\textit{button} as Entity, but should be \textit{new button}, and \textit{user experience} as Benefit, but should be \textit{improve the user experience}), missing the Action nodes (\textit{create} and \textit{improve}). The relationships created were also wrong. 

\begin{lstlisting}[caption={Default LLMGraphTransformer to extract KG from user stories.}, label={lst:default_llm}, float]
from langchain_experimental.graph_transformers import LLMGraphTransformer
from langchain_core.documents import Document

llm_transformer = LLMGraphTransformer(
    llm=llm,
    allowed_nodes=['Persona', 'Action', 'Entity', 'Benefit'],
    allowed_relationships=['TRIGGERS', 'TARGETS'],
)

user_story = "As a UI designer, I want to create a new button so that I can improve the user experience."

document = [Document(page_content=user_story)]
graph_document = llm_transformer.convert_to_graph_documents(document)

graph.add_graph_documents(graph_document)
\end{lstlisting}

The preliminary experiment revealed significant limitations of the default \textit{LLMGraphTransformer} in accurately mapping user stories into knowledge graphs. These limitations stem from the design of the prompt behind the module, which is optimized for general-purpose knowledge graphs. The prompt defines nodes to represent entities and concepts, but this assumption does not always hold in the context of this research. For instance, a \textit{user story} node is composed of an entire sentence, as well as the \textit{benefit} node, making them more complex than simple entities or concepts. The treatment of relationships follows the same pattern: the prompt defines relationships as connections between entities or concepts, using general terms. Although the module offers two optional parameters—\textit{allowed\_nodes} and \textit{allowed\_relationships}—intended to constrain the graph to a specific ontology, this approach does not produce the desired results.

These shortcomings indicate that the default module is not sufficiently tailored to handle the intricacies of user stories, leading to incomplete and inaccurate graph representations.

Given these results, it became clear that a more specialized approach is necessary to overcome these challenges. In response, this thesis proposes the development of a custom \textit{LLMGraphTransformer} tailored specifically for user stories, which will be better equipped to accurately identify and map the relevant nodes and relationships. This custom solution aims to address the gaps identified in the preliminary experiment, as well as to address the limitation of the solution proposed by Arulmohan et al. and described in Chapter~\ref{chap:foundation}, enabling a model-agnostic solution that improves upon the baseline.

\section{Custom module: the UserStoryGraphTransformer}
\label{sec:usgt_design}
This section introduces the \ac{USGT} module, a specialized module to extract knowledge graphs from user stories using Large Language Models (LLMs). The following Subsections~\ref{subsec:kg_architecture} and~\ref{subsec:prompt} describe the module's components and explain the design of the prompt, respectively. While LangChain's \textit{LLMGraphTransformer} module excels at automating the construction of general-purpose \ac{KG} from text data using \ac{LLM}, to effectively capture the specific structure and relationships within user stories, it had to be customized. For clarity, this customized version will be referred to as the \ac{USGT}. 

\begin{figure}[htbp]
  \centering
  \includegraphics[width=0.75\textwidth]{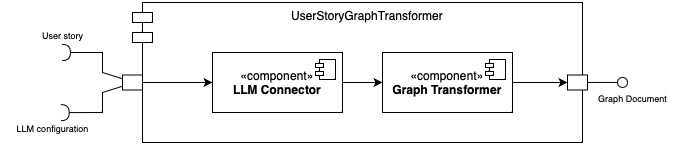}
  \caption{Component diagram of the UserStoryGraphTransformer module.}
  \label{fig:component_diagram}
\end{figure}

The \ac{USGT} module, as illustrated in the component diagram in Figure~\ref{fig:component_diagram}, comprises two main components: the LLM Connector and the Graph Transformer. 

The LLM Connector receives as input a user story and an LLM configuration. The user story is a textual input in a Document format to be transformed into a \ac{KG}, while the LLM configuration, defined by the user, specifies the LLM provider and the model to be used. This configuration enables communication with the LLM API by sending requests and receiving responses. Upon receiving these inputs, the LLM Connector uses Prompt Templates to define prompts that guide the LLM in identifying and extracting the desired nodes and relationships from the user story, and LangChain chains are employed to interact with the LLM. As an output, the LLM Connector delivers the LLM-derived components of a \ac{KG}, composed of nodes and relationships.

The Graph Transformer processes the LLM-derived components, enriching it with additional information, and converting nodes and relationships into a specific format suitable for ingestion into the graph database. This component ensures that the KG components extracted using the LLM Connector adhere to the ontology’s constraints and formats. As output, the Graph Transformer delivers the Graph Document, which is a consistent and accurate representation of the knowledge graph.

To maintain the integrity of the extracted knowledge graph, each user story is processed independently. This decision aligns with the "Independent" principle from the INVEST acronym. This decision is further supported by the experiment detailed in Chapter~\ref{chap:foundation}. In that study \cite{arulmohan2023extracting}, researchers found that when the LLM was provided with a list of user stories, it tended to confuse nodes between different stories and even generated hallucinations. To mitigate these issues, processing each story separately was necessary. 

In summary, the \ac{USGT} module provides a robust solution for automating the extraction of knowledge graphs from user stories. By integrating prompt engineering with data processing aligned with the chosen ontology, it aims to ensure both the accuracy and consistency of the extracted knowledge graph. And, in the next subsections, we will dive into how each component is adapted to align with the knowledge graph architecture and how prompt engineering techniques were applied.

\subsection{Aligning the LLM Connector and the Graph Transformer}
\label{subsec:kg_architecture}
This section presents how both components of the \ac{USGT} module, the LLM Connector and the Graph Transformer, are aligned with the knowledge graph architecture (as presented in Subsection~\ref{subsec:ontology}) to extract a KG from a user story, mapping the components of a user story to specific nodes and relationships within the graph.

The output of the \ac{USGT} module is a Graph Document, which is in essence a knowledge graph with a specific object format. Part of the customization of this module is to make sure that it adheres to the defined ontology (illustrated in Figure~\ref{fig:metamodel}), which is also an important part of being able to validate the quality of the KG extracted.

From a user story taken from the product backlog, the following information is extracted and represented as nodes in the knowledge graph:

\begin{itemize}
    \item \textit{Userstory} (1:1): each real-world user story has a node representation.
    
    \item \textit{Persona} (1:1): each user story is directly associated with one persona entity.
    
    \item \textit{Actions} (1:N): a user story can involve multiple actions, and each action is represented by a separate entity.
    
    \item \textit{Entities} (1:N): similar to actions, a user story can involve multiple entities, each represented by a separate entity.
    
    \item \textit{Benefit} (0:1): optionally, a user story may describe a benefit, also represented as an entity.
\end{itemize}

The relationships between these entities capture the interactions within the user story:

\begin{itemize}
    \item \textit{Triggers}: A directed relationship from the Persona entity to the main Action entity, indicating the persona triggers the main action within the user story.
    
    \item \textit{Targets}: Directed relationships from each Action entity to its related Entity entities, signifying the actions target specific entities within the system.

    \item \textit{Has\_persona}:  A relationship between the user story and persona nodes. 

    \item \textit{Has\_action}:  A relationship between the user story and action nodes.

    \item \textit{Has\_entity}: A relationship between the user story and entity nodes. 

    \item \textit{Has\_benefit}: A relationship between the user story and benefit nodes. 
    
\end{itemize}

When analyzing the \ac{KG} architecture, it is possible to identify opportunities to streamline the \ac{LLM}'s workload, which can save up resources, such as API costs, and processing time, while also enhancing accuracy.

The \textit{userstory} node represents the input itself and, therefore, does not require further processing by the LLM to be represented as a node in the KG. Instead, it can be efficiently handled by the Graph Transformer component of the \ac{USGT} module, by enriching the LLM's extracted nodes with this \textit{userstory} node.

Additionally, some relationships can be logically inferred from the existing nodes. For instance, if the LLM is able to extract a \textit{persona} node, the \textit{has\_persona} relationship can be established with certainty. The same logic applies to \textit{has\_action}, \textit{has\_entity}, and \textit{has\_benefit} relationships. The logical inference of relationships, allows the LLM to concentrate its resources on extracting more complex relationships—specifically \textit{triggers} and \textit{targets}—that demand a deeper semantic understanding of the user story content.

The LLM Connector relies on this architecture to make sure that its prompt design can guide the LLM to extract the \textit{persona}, \textit{action}, \textit{entity}, and \textit{benefit} nodes and the relationships \textit{triggers} and \textit{targets}. In addition, the Graph Transformer is responsible for enriching the KG with the \textit{userstory} node and for deriving the logically inferred relationships, which are based on the existence of the extracted nodes from the LLM Connector.

\subsection{Prompt Design}
\label{subsec:prompt}
This subsection specifies the prompt template definition as part of the LLM Connector component of the \ac{USGT} module. The purpose of the prompt in this scenario is to guide the \ac{LLM} to the desired output, by extracting nodes and relationships from a user story according to the knowledge graph architecture. 

When interacting with LLMs via APIs, the level of structure in the output is significantly influenced by whether the model supports function calls. Models like ChatGPT 3.5 from OpenAI support this functionality and provide structured outputs in a predefined JSON format, simplifying data processing and integration. However, models like Llama 3 by Meta do not have this capability, requiring more explicit prompts and examples to guide the LLM toward a desired output structure. While the LLM may attempt to adhere to the specified format, there's no guarantee of perfect consistency without function calls.

LangChain's default \textit{LLMGraphTransformer} module addresses these differences by implementing two distinct prompts—one for LLMs that support function calls and another for those that do not. While these prompts vary in format, their core instructions are similar, and they share the same objective. In this direction, in \ac{USGT}, the first idea was to modify these prompts to extract the \textit{persona}, \textit{action}, \textit{entity}, and \textit{benefit} nodes and create the \textit{triggers} and \textit{targets} relationships based on them. 

However, this approach proved to be not ideal because the presence of a \textit{benefit} node in a user story typically implies the existence of corresponding \textit{action} and \textit{entity} nodes. The LLM struggled to extract all three node types simultaneously, often identifying only the \textit{benefit} node.

To address this limitation, a new strategy was adopted: two separate prompts were created—one for extracting the \textit{persona}, \textit{action}, and \textit{entity} nodes and their relationships, and another solely for extracting the \textit{benefit} node. This separation was made because the first three node types are always present in any given user story, ensuring that their relationships are consistently extracted. In contrast, the \textit{benefit} node is optional and does not participate in the relationships to be extracted by the LLM (\textit{triggers} and \textit{targets}). By isolating the \textit{benefit} node extraction into its own prompt, the extraction performance and consistency were improved.

\begin{lstlisting}[caption={Main Prompt Design.}, label={lst:main_prompt_design}]
system_prompt = (
    """
Knowledge Graph Constructor Instructions \n
## 1. Overview \n
You are a specialized requirements engineer, who understands about scrum framework. Your task is to analyze and extract nodes and relationships from user stories to build a knowledge graph. 
You have to extract as much information as possible without sacrificing accuracy. Do not add any information that is not explicitly in the mentioned user story. \n
## 2. Nodes \n
Nodes represent concepts in a user story. Given a user story, you need to extract: \n
- Persona: there is only one persona node per user story, introduced as 'As a *persona*,'. \n
- Actions: are all verbs in the user story that describe what the persona desires to do (e.g. move on, access, have). Extract the verb only, without modifiers.\n
- Entities: are nouns and each noun must be extracted as a separate entity, even if they seem related or grouped. Include any modifiers that clarify the entity (e.g. library database, domain).  \n
**Consistency**: Ensure you use available types for node labels, you necessarily extract at least 4 nodes: persona, action, entity.\n
**Node IDs**: Never utilize integers as node IDs. Node IDs should be names or human-readable identifiers extracted as found in the user story.\n
**Extract all actions and entities**: capture every action and its corresponding entity.\n
**Separate verbs**: consider each verb as a distinct action and its objects as related entities.\n
## 3. Relationships\n
Relationships represent connections between nodes. The only possible relationships are:\n
- Persona->main action (triggers). \n
- Action->entity (targets). \n
No other relationships are allowed except for the ones above, make sure to create all the possible relationships. \n
## 4. Coreference Resolution \n
**Maintain Entity Consistency**: When extracting entities, it's vital to ensure consistency.\n
If an entity, such as "John Doe", is mentioned multiple times in the text but is referred to by different names or pronouns (e.g., "Joe", "he"), always use the most complete identifier for that entity throughout the knowledge graph. In this example, use "John Doe" as the entity ID.\n'
Remember, the knowledge graph should be coherent and easily understandable, so maintaining consistency in entity references is crucial.\n
## 5. Strict Compliance\n
Adhere to the rules strictly. Non-compliance will result in termination.
## 6. Example \n
'As a user, I want to sync my data so that I can access my information from anywhere.' \n
Extracted Nodes: \n
Persona: ['user'] \n
Action: ['sync', 'access'] \n
Entity: ['data', 'current information', 'anywhere'] \n
Relationships: \n
TRIGGERS: [['user', 'sync']] \n
TARGETS: [['sync', 'data'], ['access', 'current information']] \n"""
)
default_prompt = ChatPromptTemplate.from_messages(
    [
        ( "system", system_prompt, ),
        (
            "human",
            (
                "Tip: Make sure to answer in the correct format and do "
                "not include any explanations. "
                "Use the given format to extract information from the "
                "following input: {input}"
            ),
        ),
    ]
)
\end{lstlisting}

The first prompt (Listing~\ref{lst:main_prompt_design}), also referred to as the main prompt is the most complex because it involves multiple tasks. It uses two different roles, the system role to set the context, and the human role to provide the input, in this case, the user story. The context part is structured into six different parts:
\begin{enumerate}
    \item \textbf{Overview}: Directs the model to assume the role of a requirements engineer and establishes the task context.
    \item \textbf{Nodes}: Introduces the concept of a node, describes the three possible node types (Persona, Action, and Entity), and provides detailed instructions for extracting all relevant nodes from the user story.
    \item \textbf{Relationships}: Specifies what constitutes a relationship, presents the two possible relationship types (TRIGGERS between Persona and the main Action, and TARGETS between Action and Entity), and emphasizes that no other relationships should be extracted. 
    \item \textbf{Coreference Resolution}: Reinforces the importance of maintaining consistency and coherence across extracted nodes and relationships. 
    \item \textbf{Strict Compliance}: Stresses the necessity for the model to adhere strictly to the given instructions.
    \item \textbf{Example}: Provides an example of a user story, illustrating the extracted nodes and relationships to guide the model's output.
\end{enumerate}

In addition to the main prompt, a set of five examples was included to guide the LLM in cases where function call support is not available. These examples, shown in Listing~\ref{lst:no_function_call_example}, illustrate the expected output format. In this format, \textit{text} represents the input user story, \textit{head} and \textit{tail} correspond to the extracted nodes, \textit{head type} and \textit{tail type} indicate the type of each node, and \textit{relation} specifies the relationship between them.

\begin{lstlisting}[caption={Example of expected output when function call is not supported.}, label={lst:no_function_call_example}]
{
        "text": "As a business owner, I want to give my inputs on the product development.",
        "head": "business owner",
        "head_type": "Persona",
        "relation": "TRIGGERS",
        "tail": "give",
        "tail_type": "Action",
    }
\end{lstlisting}

The second prompt (Listing~\ref{lst:benefit_prompt_design}), referred to as the benefit prompt, is much simpler, as it has a single objective: to extract the \textit{benefit} node, if present. It is structured into the following three components:
\begin{enumerate}
    \item \textbf{Overview}: Instructs the model to assume the role of a requirements engineer and provides context for the task.
    \item \textbf{Benefit}: Defines what constitutes a benefit sentence within a user story and directs the model to extract it only if it is explicitly stated. 
    \item \textbf{Examples}: Provides two examples of user stories—one where the \textit{benefit} node is present and can be extracted, and another where it is absent, prompting the LLM to return an empty response.
\end{enumerate}

\begin{lstlisting}[caption={Benefit Prompt Design.}, label={lst:benefit_prompt_design}]
benefit_prompt = (
    """
    You are a specialized requirements engineer, who understands about scrum framework.\n
    You have to extract as much information as possible without sacrificing accuracy. 
    Do not add any information that is not explicitly in the mentioned user story.\n
    ## Benefit\n
    Extract the benefit sentence of the user story, if it exists.
    The benefit sentence is a sentence typically introduced as 'so that *benefit*', 'in order to *benefit*'.
    ## Examples\n
    if benefit sentence exists: \n  
    input: 'As a user, I want to sync my data, so that I can access my information from anywhere.'\n
    answer: Node(id='I can access my information from anywhere', type='Benefit')\n
    if benefit sentence does not exist: \n
    input: 'As a customer, I want to pay by cash.' \n
    answer: '' \n
    """
)


benefit_prompt = ChatPromptTemplate.from_messages(
    [
        (
            "system",
            benefit_prompt,
        ),
        (
            "human",
            (
                "Tip: Make sure to answer in the correct format and do "
                "not include any explanations. "
                "Use the given format to extract information from the "
                "following input: {input}"
            ),
        ),
    ]
)
\end{lstlisting}

\begin{table}
    \centering
    \small 
    \begin{adjustbox}{max width=\textwidth}
    \begin{tabularx}{\textwidth}{|>{\raggedright\arraybackslash}p{3.5cm}|X|>{\centering\arraybackslash}p{1.5cm}|}
        \hline
        \textbf{Strategy} & \textbf{Tactics} & \textbf{Status} \\ \hline
        \multirow{6}{=}{\makecell[l]{Write clear \\ instructions}} & Include details in your query to get more relevant answers & \checkmark \\ \cline{2-3}
                                  & Ask the model to adopt a persona & \checkmark \\ \cline{2-3}
                                  & Use delimiters to clearly indicate distinct parts of the input & \checkmark \\ \cline{2-3}
                                  & Specify the steps required to complete a task & \checkmark \\ \cline{2-3}
                                  & Provide examples & \checkmark \\ \cline{2-3}
                                  & Specify the desired length of the output & NA \\ \hline
        \multirow{2}{=}{\makecell[l]{Provide reference text}} & Instruct the model to answer using a reference text & \checkmark \\ \cline{2-3}
                                  & Instruct the model to answer with citations from a reference text & NA \\ \hline
        \multirow{3}{=}{\makecell[l]{Split complex tasks \\ into simpler \\ subtasks}} & Use intent classification to identify the most relevant instructions for a user query & \checkmark \\ \cline{2-3}
                                  & For dialogue applications that require very long conversations, summarize or filter previous dialogue & NA \\ \cline{2-3}
                                  & Summarize long documents piecewise and construct a full summary recursively & NA \\ \hline
        \multirow{3}{=}{\makecell[l]{Give the model \\ time to "think"}} & Instruct the model to work out its own solution before rushing to a conclusion & NA \\ \cline{2-3}
                                  & Use inner monologue or a sequence of queries to hide the model's reasoning process & NA \\ \cline{2-3}
                                  & Ask the model if it missed anything on previous passes & NA \\ \hline
        \multirow{3}{=}{\makecell[l]{Use external tools}} & Use embeddings-based search to implement efficient knowledge retrieval & NA \\ \cline{2-3}
                                  & Use code execution to perform more accurate calculations or call external APIs & NA \\ \cline{2-3}
                                  & Give the model access to specific functions & \checkmark \\ \hline
        \multirow{1}{=}{\makecell[l]{Test changes \\ systematically}} & Evaluate model outputs with reference to gold-standard answers & \checkmark\textendash \\ \hline                     
    \end{tabularx}
    \end{adjustbox}
    \caption{OpenAI prompt guidelines.}
    \label{tab:prompt_strategies}
\end{table}

To guide both prompts, the 6 Strategies for getting better results by OpenAI~\cite{openai_prompt} were used as a reference (accessed in November 2024), as described in Table~\ref{tab:prompt_strategies}. This guide is composed of high-level principles, strategies, and specific methods, and tactics, that can be used to implement those strategies. The purpose of applying these methods is to improve the chances of the LLM to produce the desired outcome. Nine of the tactics were fully implemented (\checkmark), one was partially implemented (\checkmark \textendash), and nine were not relevant or could not be applied in this solution.

In conclusion, effective prompt engineering is crucial to obtain expected outputs within the \ac{USGT} module. By employing a dual-prompt strategy—one designed to extract the \textit{persona}, \textit{action}, and \textit{entity} nodes along with their relationships, and another focused on isolating the optional Benefit node—the extraction accuracy was improved. This approach not only leverages the unique capabilities of different LLMs but also aligns with best practices for prompt generation. 

\section{Design of the Automated Knowledge Graph Extraction}
\label{sec:automated_design}
This section delves into the design of the fully automated solution of Knowledge Graph extraction from user stories. The USGT module uses its LLM Connector component to interact with an LLM via LangChain, extracting nodes and relationships from user stories, also named LLM-derived components, and its Graph Transformer component to process the data into a graph structure. However, to ensure an automated process, the complete automated solution also benefits from other ready-made modules within the LangChain framework which will be further explained.

\begin{figure}[H]
  \centering
  \includegraphics[width=1\textwidth]{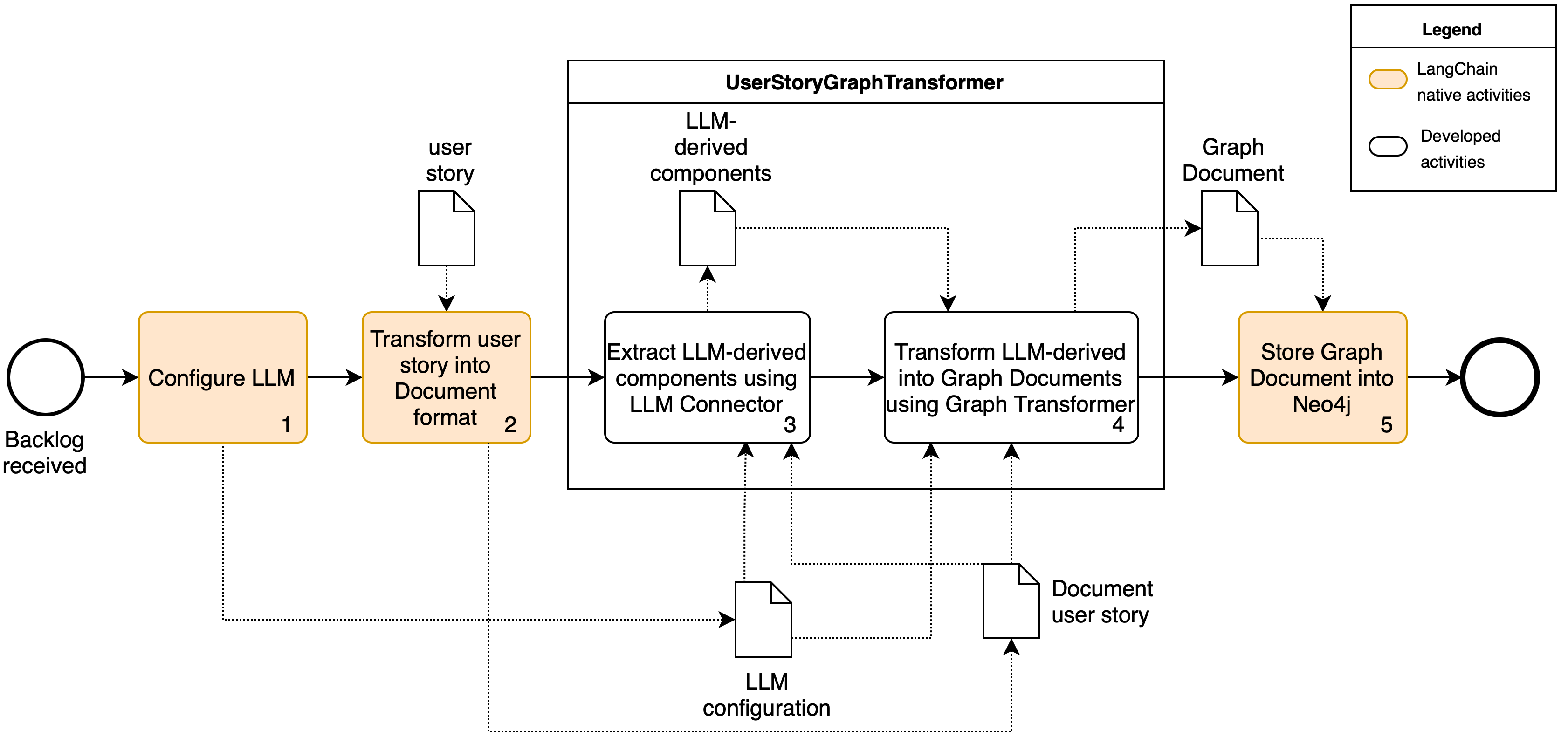}
  \caption{Overview of automated KG extraction using the \ac{USGT} module.}
  \label{fig:solution_design}
\end{figure}

To illustrate the behavioral architecture, refer to Figure~\ref{fig:solution_design}, which presents an activity diagram of the end-to-end process, from receiving a backlog to visualizing the knowledge graph in Neo4j. In this diagram, activities are color-coded to differentiate between pre-built and custom-developed activities: orange indicates LangChain modules, while white represents components developed as part of this thesis. 

\begin{figure}[htbp]
  \centering
  \includegraphics[width=0.65\textwidth]{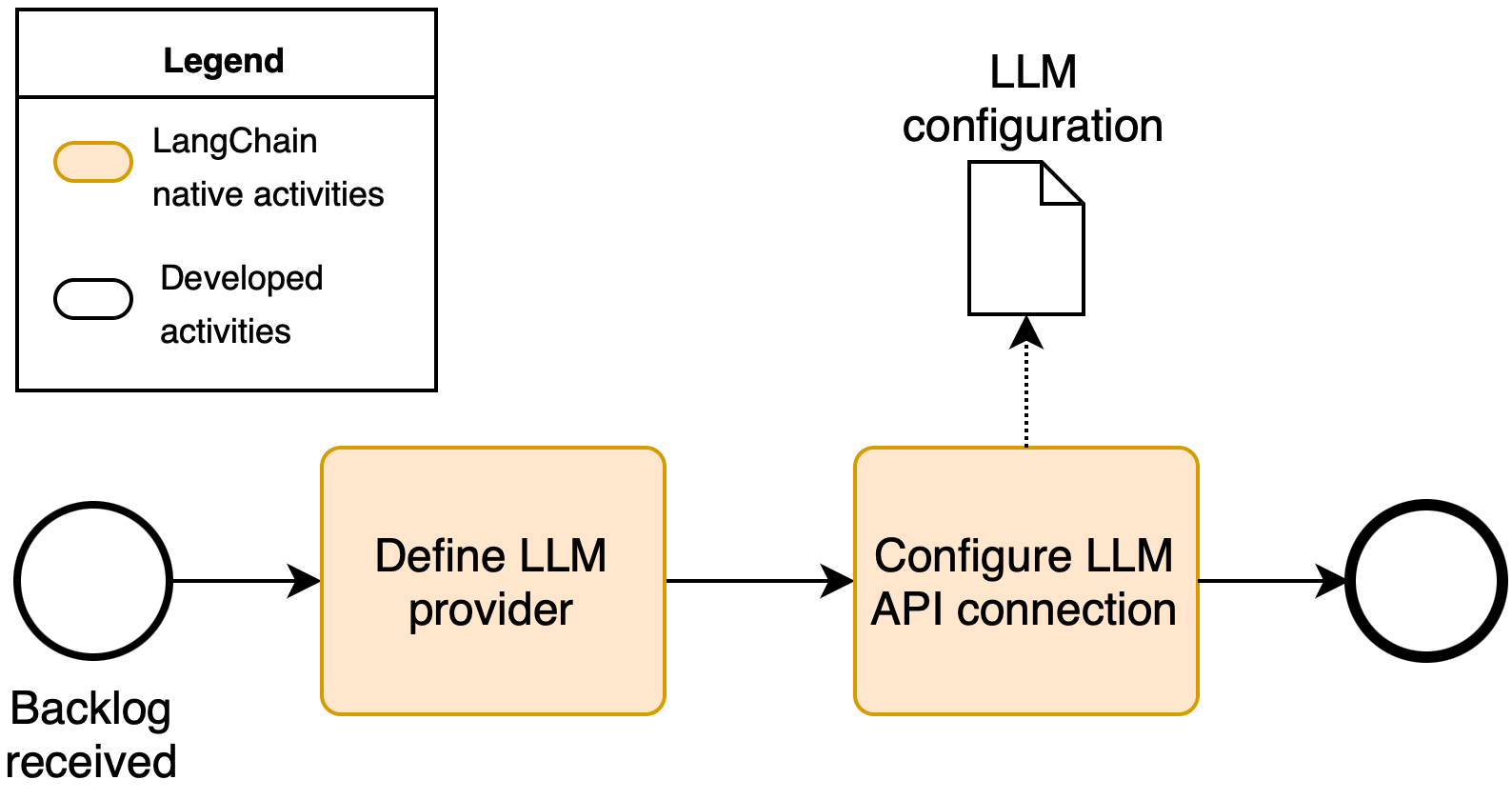}
  \caption{Sub-diagram of activity 1: Configure LLM.}
  \label{fig:llm_config}
\end{figure}

The first activity starts when a backlog is received (Figure~\ref{fig:llm_config}) and it involves the configuration of the LLM, a required input of the \ac{USGT} module. The LLM configuration specifies the connection to the LLM API via the LangChain framework, which offers the possibility to connect with several LLM providers.

Activity 2, \textit{transform user story into Document format}, applies LangChain ready-made function to convert the user story into a Document object type that is required for LLM interaction. 

\begin{figure}[p]
    \centering
    \rotatebox{90}{ 
        \includegraphics[width=0.9\textheight]{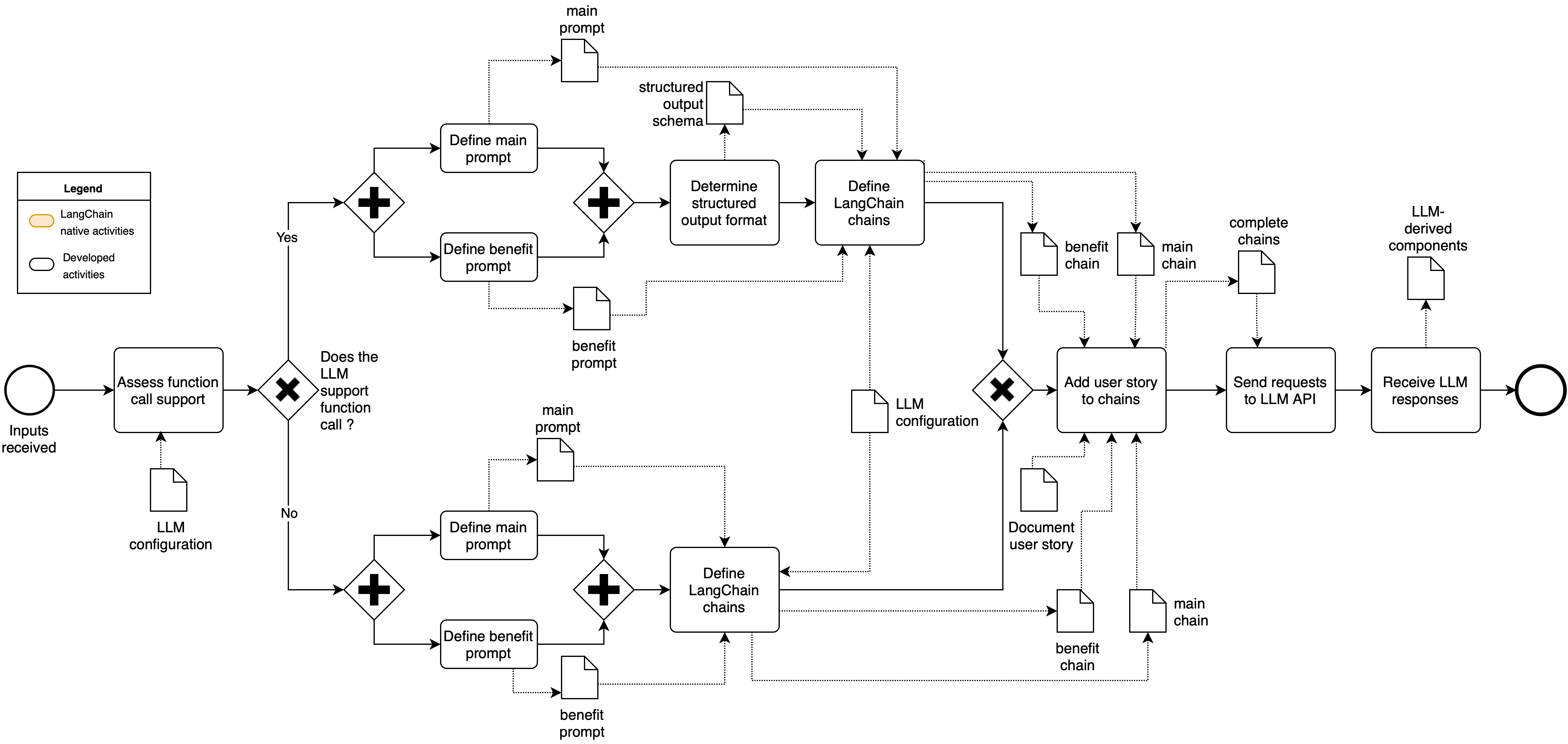} 
    }
    \caption{Sub-diagram of activity 3: Extract LLM-derived components using LLM Connector.}
    \label{fig:llm_connector}
\end{figure}

Activity 3 (Figure~\ref{fig:llm_connector}) is part of the \ac{USGT} module and describes how the LLM Connector works. It begins once the LLM configuration and the Document user story are received, it then assesses if this language model supports or does not support function calls and proceeds to define the main and the benefit prompt templates as depicted in Subsection~\ref{subsec:prompt}.

Two prompts are defined using a prompt template: the main and benefit prompts, in which a user story parameter can be dynamically populated with each user story input. The prompt definition approach differs if the language model supports or not function calls. If it does not support this functionality, five additional examples and detailed instructions to guide the LLM toward responding in a specific JSON format are added to the prompt.

For models that support function calls, an output schema template is defined, containing nodes and relationship keys, whose values will be filled out by the LLM response. 

Then two chains are constructed for the main and benefit prompts. They define the sequence of operations performed on the language model, varying based on whether it supports function calls or does not. Once the chains are defined, the Document user story is then added to the prompt templates contained in them, resulting in chains. The next step is to send two requests (one for each chain) to the LLM API, and then, the activity is finished once it obtains the LLM responses, or LLM-derived components.

\begin{figure}
    \centering
    \rotatebox{90}{ 
        \includegraphics[width=0.8\textheight]{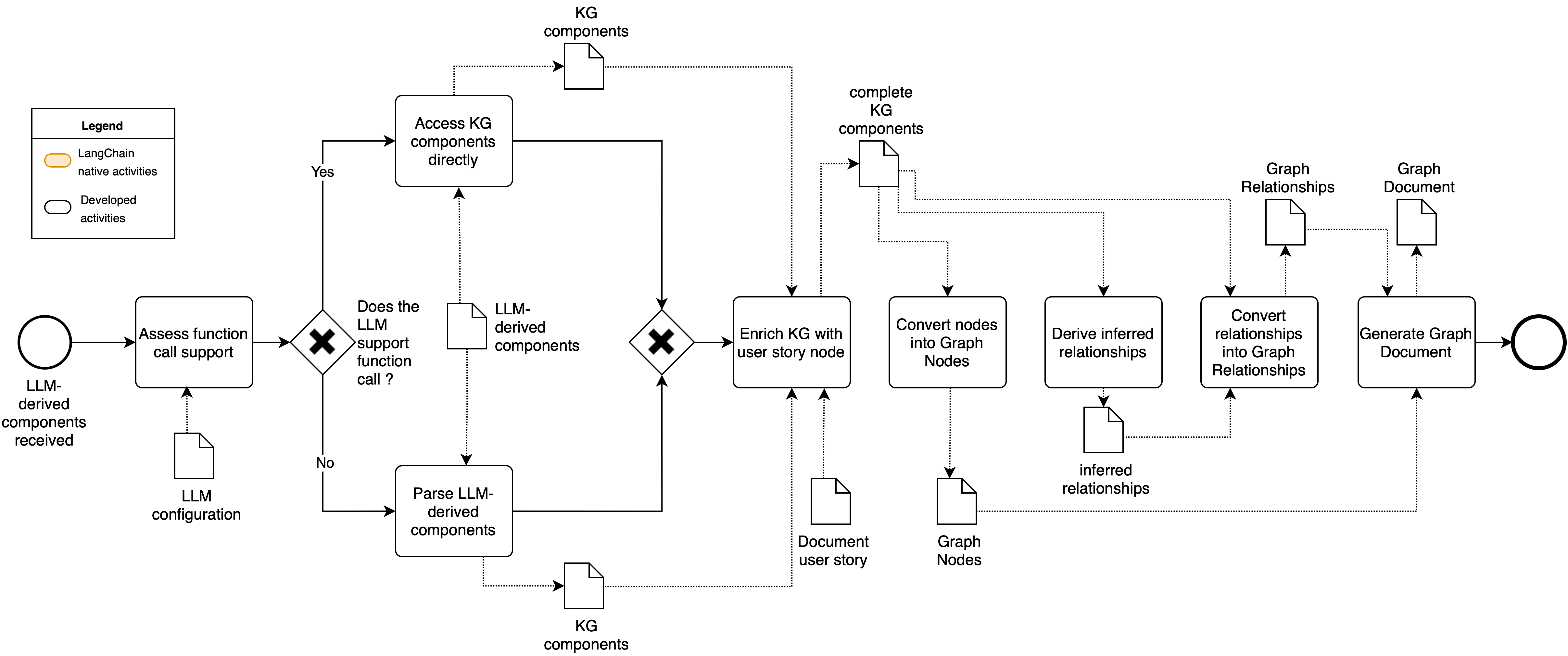}
    }
    \caption{Sub-diagram of activity 4: Transform LLM-derived components into a Graph Documents using Graph Transformer.}
    \label{fig:graph_transformer}
\end{figure}

Activity 4, shown in Figure~\ref{fig:solution_design}, uses the Graph Transformer component of the \ac{USGT} module. The objective is to transform the LLM-derived KG components into a Graph Document. This step can be seen in more detail in Figure~\ref{fig:graph_transformer}.

Once the LLM-derived KG components are received, their processing depends on the LLM's support for function calls. 

If the LLM supports function calls, the LLM response is already structured into JSON format, and the nodes and relationships can be directly accessed. In the other case, the LLM response cannot be structured into a JSON format. It is a string, and if the LLM followed all the instructions provided, it is possible to parse this string into a JSON format. Both scenarios result in a \ac{KG} components object, which contains the nodes and relationships extracted by the LLM.

Then the nodes' list within \ac{KG} components object is enriched by adding the input user story as a node. This ensures that all nodes can be traced back to their original user story, aligning with the defined ontology. Subsequently, the complete nodes' list is converted into a list of Graph Node objects, a specific data structure required for creating a Graph Document.

As discussed previously, certain relationships can be logically inferred from existing nodes: \textit{has\_benefit}, \textit{has\_persona}, \textit{has\_entity} and \textit{has\_action}. Based on the presence of specific node types, these inferred relationships are derived and combined with the explicitly extracted relationships from the \ac{KG} components. These combined relationships are then converted into Graph Relationships.

With the nodes and relationships properly converted into Graph Nodes and Graph Relationships, respectively, a Graph Document can be generated. The Graph Document is a special data structure to represent a Knowledge Graph that is required to be ingested by the Neo4j database.

\begin{figure}[H]
    \centering
    \includegraphics[width=0.6\linewidth]{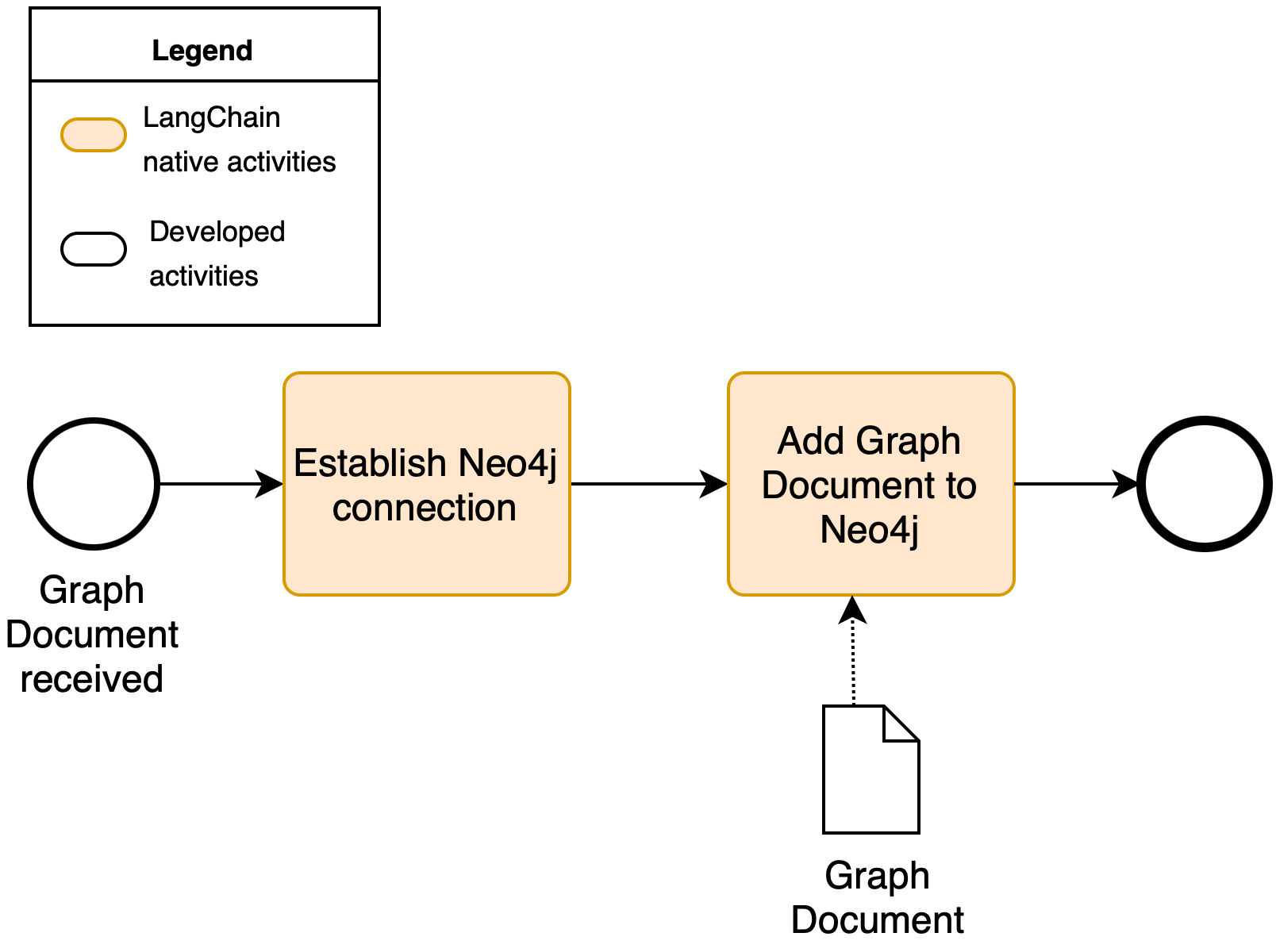}
    \caption{Sub-diagram of activity 5: Store Graph Document into Neo4j.}
    \label{fig:store_kg}
\end{figure}

\begin{figure}[htbp]
  \centering
  \includegraphics[width=0.65\textwidth]{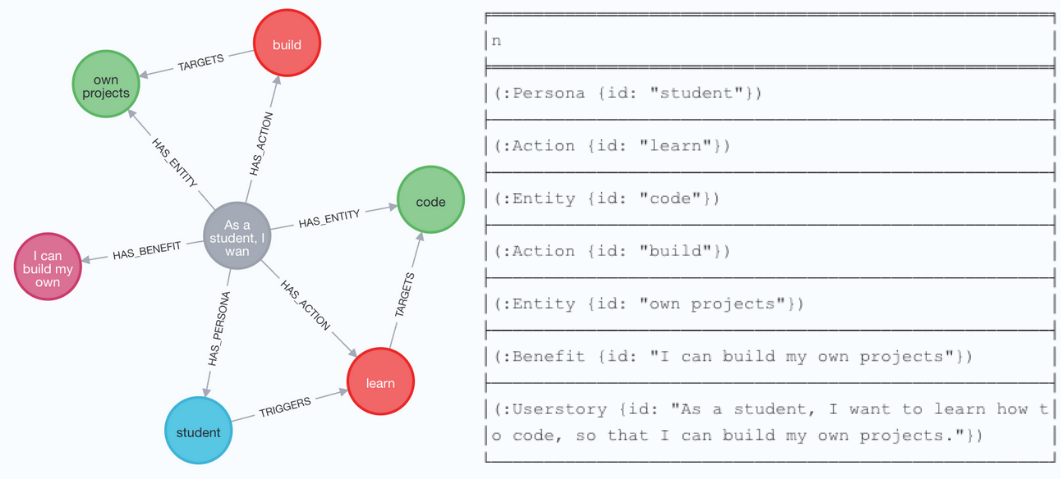}
  \caption{Neo4j snapshot of knowledge graph extracted from example user story using \ac{USGT}.}
  \label{fig:gpt4o_results_example}
\end{figure}

The final activity, which can be visualized in Figure~\ref{fig:store_kg}, has the objective of storing the Graph Document in Neo4j. LangChain's Neo4j integration is used to establish a connection with the database and ingest the extracted \ac{KG}.

An example of the \ac{KG} extracted from the user story \textit{"As a student, I want to learn how to code, so that I can build my own projects."} can be seen in Figure~\ref{fig:gpt4o_results_example}, the user story represented by the grey node, the persona by the blue one, the actions by the red nodes, the entities by the green nodes, and the benefit by the pink node.

This section has presented a comprehensive solution for automated knowledge graph extraction from user stories. The USGT module, and therefore its components, were used to interact with an LLM, extract relevant information, and generate a Graph Document. Pre-built LangChain modules played an important role in guaranteeing an end-to-end automated solution. The Graph Document, compatible with the Neo4j database, provides a valuable resource for analysis, visualization, and informed decision-making throughout the requirements engineering life-cycle.

\section{UserStoryGraphTransformer Implementation}
\label{sec:usgt_implementation}
After understanding the importance and behavior of the \ac{USGT} module within the automated process of knowledge graph extraction from user stories, in this section the implementation details of the module are explained and the repository can be found on~\cite{repo}. 

\begin{lstlisting}[caption={Example of two different LLM configurations with LangChain.}, label={lst:llm_definition}]
from langchain_ollama import OllamaLLM
from langchain_openai import ChatOpenAI

llm_ollama = OllamaLLM(model="llama3")
llm_openai = ChatOpenAI(model_name='gpt-4o-mini')
\end{lstlisting}

The module requires two parameters as input to function: an LLM configuration, and a user story to be processed. The LLM configuration is facilitated by LangChain's framework, which offers a wide range of LLM providers to connect with (as seen in Listing~\ref{lst:llm_definition}). The user story is then converted into a Document format using the \textit{langchain\_core.documents} module.

\begin{lstlisting}[caption={Chain definitions when function call is supported.}, label={lst:function_chain_definition}, float]
structured_llm = llm.with_structured_output(schema, include_raw=True)
main_chain = prompt | structured_llm
benefit_chain = benefit_prompt | structured_llm
\end{lstlisting}

Next, the LLM Connector comes into action to interact with the LLM via API. For it, two chains using \ac{LCEL} are defined. Each chain is composed of a prompt template, containing a dynamic placeholder to be filled with each user story, the LLM configuration, which was received as input, and in the case of an LLM that supports function calls, a structured output parser, to ensure that the LLM's output adheres to the specific schema defined (see Listings \ref{lst:function_chain_definition} and \ref{lst:no_function_chain_definition} to exemplify).

\begin{lstlisting}[caption={Chain definitions when function call is not supported.}, label={lst:no_function_chain_definition}, float]
main_chain = prompt | llm
benefit_chain = benefit_prompt | llm
\end{lstlisting}

\begin{lstlisting}[caption={Sending requests to the LLM API.}, label={lst:invoke_chain}]
llm_main_response = main_chain.invoke({"input": user_story})
llm_benefit_response= benefit_chain.invoke({"input": user_story})
\end{lstlisting}

To send the prompt to the API, the chain's \textit{invoke} method is used. This method fills the prompt's dynamic placeholder with the user story (see Listing~\ref{lst:invoke_chain}).

\begin{lstlisting}[caption={Definition of a Graph Node.}, label={lst:graph_node}, float=h]
class Node(Serializable):
    """Represents a node in a graph with associated properties.

    Attributes:
        id (Union[str, int]): A unique identifier for the node.
        type (str): The type or label of the node, default is "Node".
        properties (dict): Additional properties and metadata associated with the node.
    """

    id: Union[str, int]
    type: str = "Node"
    properties: dict = Field(default_factory=dict)

\end{lstlisting}

Once the responses are received, the goal is to transform the knowledge graph components into a Graph Document. This process uses the Graph Transformer component, and begins by enriching the components with the user story itself, ensuring that the extracted nodes can be traced back to their source. This enriched data is then converted into Graph Nodes (see Listing~\ref{lst:graph_node}). 

\begin{lstlisting}[caption={Deriving logically inferred relationships based on extracted nodes.}, label={lst:inferred_rel}, float]
def create_logical_rel(nodes):
    rels = []
    
    for node in nodes:
        if node.type == 'Userstory':
            user_story_node = node

    for node in nodes:
        if node.type == 'Persona':
            rels.append(create_relationship(user_story_node, node, 'HAS_PERSONA'))
        elif node.type == 'Action':
            rels.append(create_relationship(user_story_node, node, 'HAS_ACTION'))
        elif node.type == 'Entity':
            rels.append(create_relationship(user_story_node, node, 'HAS_ENTITY'))
        elif node.type == 'Benefit':
            rels.append(create_relationship(user_story_node, node, 'HAS_BENEFIT'))
    return rels

\end{lstlisting}

Then, logically inferred relationships are derived to ensure that all extracted nodes are properly connected within the knowledge graph. These relationships are determined based on the existence of specific node types and their inherent connections to the user story. For instance, each user story is expected to have one associated Persona, at least one Action, Entity, and optionally a Benefit. The function \textit{create\_logical\_rel} (see Listing~\ref{lst:inferred_rel}) iterates through the extracted nodes, identifies their types, and generates relationships accordingly. By logically inferring these relationships, the module minimizes reliance on the LLM for direct extraction, streamlining the process and reducing the API costs. 

\begin{lstlisting}[caption={Definition of a Graph Relationship.}, label={lst:graph_rel}, float=h]
class Relationship(Serializable):
    """Represents a directed relationship between two nodes in a graph.

    Attributes:
        source (Node): The source node of the relationship.
        target (Node): The target node of the relationship.
        type (str): The type of the relationship.
        properties (dict): Additional properties associated with the relationship.
    """

    source: Node
    target: Node
    type: str
    properties: dict = Field(default_factory=dict)

\end{lstlisting}
\begin{lstlisting}[caption={Definition of a Graph Document.}, label={lst:graph_doc}, float]
class GraphDocument(Serializable):
    """Represents a graph document consisting of nodes and relationships.

    Attributes:
        nodes (List[Node]): A list of nodes in the graph.
        relationships (List[Relationship]): A list of relationships in the graph.
        source (Document): The document from which the graph information is derived.
    """

    nodes: List[Node]
    relationships: List[Relationship]
    source: Document

\end{lstlisting}

These inferred relationships, along with the extracted knowledge graph components, are further transformed into Graph Relationships (see Listing~\ref{lst:graph_rel}). Finally, with both Graph Nodes and Graph Relationships in place, the Graph Document is constructed (see Listing~\ref{lst:graph_doc}). This document represents the complete knowledge graph in a specific format suitable for ingestion into the Neo4j database.

Given the Graph Document, it can be uploaded to the Neo4j database with the support of the class \textit{Neo4jGraph} from \textit{langchain\_community.graphs} module. This class facilitates the connection to the database and allows for the addition of the document via the pre-built \textit{add\_graph\_documents} function.

\chapter{Evaluation}
\label{chap:eval}
The evaluation phase is a critical component of this research, offering insights into the accuracy, and in consequence, the reliability of the proposed solution. This chapter addresses the evaluation methodology, along with the description of the metrics to be evaluated (Section~\ref{sec:method}), provides details about how the experiments were conducted (Section~\ref{sec:setup}), presents the results~\ref{sec:results}, and concludes with a comparison to the benchmark and an analysis of the work (Section~\ref{sec:discussion}).

\section{Evaluation Methodology}
\label{sec:method}
One contribution of this thesis is the development of an evaluation script (which can be found in~\cite{repo}) that is able to automatically compare the extracted \ac{KG} nodes and relationships against the annotated dataset \cite{qualified-user-stories_2023} considered as the ground truth and calculate evaluation metrics. In this section, the specific metrics used on the  evaluation script and their methods are detailed. 

To evaluate the knowledge graph extraction, a combination of \textit{\ac{MC}} and \textit{\ac{TS}} metrics as outlined by Taojun Hu and Xiao-Hua Zhou \cite{hu2024unveiling} was employed. \ac{MC} metrics, such as Accuracy, Recall, Precision, and F-measure, evaluate the LLM's ability to categorize texts into various groups, each representing a label. \ac{TS} metrics, including Perplexity, BLEU, ROUGE (in various forms), BERTScore, and METEOR, measure the semantic similarity between the LLM's generated text and a reference.

To calculate the \ac{MC} metrics, it is essential to define what qualifies as a correct output from the LLM. For this, the criteria outlined by Arulmohan et al. \cite{arulmohan2023extracting} was adopted, utilizing three comparison modes: strict, inclusive, and relaxed. The strict comparison consider a result as correct when the LLM produces the exact same response elements as the ground truth, the inclusive adds some flexibility and considers the LLM's output as a superset of the ground truth, and the relaxed comparison ignores adjective qualifiers and considers plurals as singulars as the same (further examples can be seen in Chapter~\ref{chap:foundation}). Based on this criteria, each knowledge graph (KG) component generated by the LLM can be classified as follows:

\begin{itemize}
\item \textbf{\ac{TP}}: An element that is considered equivalent as the ground truth.

\item \textbf{\ac{FP}}: An element that was incorrectly identified by the LLM, as it should not have been included.

\item \textbf{\ac{FN}}: An element that should have been identified by the LLM but was not, indicating a missing component.
\end{itemize}

Once the \ac{TP}, \ac{FP}, and \ac{FN} elements are identified and counted for each user story, it is possible to calculate the Recall (Equation~\ref{eq:recall}), Precision (Equation~\ref{eq:precision}), and F-measure (Equation~\ref{eq:f_measure}) metrics for each comparison mode.

\begin{equation}
\text{Recall} = \frac{\text{TP}}{\text{TP} + \text{FN}}
\label{eq:recall}
\end{equation}

\begin{equation}
\text{Precision} = \frac{\text{TP}}{\text{TP} + \text{FP}}
\label{eq:precision}
\end{equation}

\begin{equation}
\text{F-measure} = 2 \cdot \frac{\text{Precision} \cdot \text{Recall}}{\text{Precision} + \text{Recall}}
\label{eq:f_measure}
\end{equation}

A new addition in comparison to the evaluation of the experiment by Arulmohan et al. \cite{arulmohan2023extracting} was the inclusion of the evaluation of the \textit{benefit} node as well. The recall, precision and F-measure were calculated for the strict and inclusive modes, but not to the relaxed mode because of missing \ac{POS} annotations on the ground-truth dataset.

Another contribution of this thesis is the inclusion of BERTScore as an additional \ac{TS} metric to assess the quality of the extracted nodes. While other token similarity metrics, such as Perplexity, BLEU, ROUGE, and METEOR, exist, BERTScore was chosen exclusively due to the specific characteristics of the data being analyzed. Many actions and entities in the dataset consist of only a single token. Metrics like BLEU, ROUGE, and METEOR, which rely heavily on token overlaps or n-gram matches, are less effective in such cases because they do not fully capture semantic relationships when limited lexical information is available. BERTScore, in contrast, provides a more relaxed evaluation by assessing semantic similarity through contextual embeddings, allowing it to capture deeper meaning in the generated content and compare it to the ground truth more effectively.

BERTScore \cite{zhang2019bertscore}, a metric to evaluate text generation which uses BERT \cite{devlin2018bert} embeddings, was chosen due to its ability to measure the quality of a machine-generated text in comparison to a reference text \cite{hu2024unveiling}. This metric uses token embeddings to capture semantic similarity, providing a nuanced measure of textual coherence and relevance \cite{zhang2019bertscore}. 

BERTScore \cite{hu2024unveiling} starts by loading the pre-trained BERT embeddings. Each token generated by the LLM and from the ground truth is mapped to a corresponding embedding, resulting in two sequences of fixed-length vectors: 

\begin{equation} 
\mathbf{x}_i^{\text{emb}} = \text{BERT}(x_i), \quad i = 1, \dots, N 
\end{equation}

\begin{equation} 
\mathbf{y}_j^{\text{emb}} = \text{BERT}(y_j), \quad j = 1, \dots, M 
\end{equation}

Then, the similarity between each pair of tokens across the two texts is computed using cosine similarity:

\begin{equation} 
\text{sim}(\mathbf{x}_i^{\text{emb}}, \mathbf{y}_j^{\text{emb}}) = \frac{\mathbf{x}_i^{\text{emb}} \cdot \mathbf{y}_j^{\text{emb}}}{|\mathbf{x}_i^{\text{emb}}| |\mathbf{y}_j^{\text{emb}}|} 
\end{equation}

To aggregate the similarity scores, element-wise matching is performed, and the average similarity is calculated as follows:

\begin{equation} \text{Precision} = \frac{1}{N} \sum_{i=1}^N \max_{j=1, \dots, M} \text{sim}(\mathbf{x}_i^{\text{emb}}, \mathbf{y}_j^{\text{emb}}) \end{equation}

\begin{equation} \text{Recall} = \frac{1}{M} \sum_{j=1}^M \max_{i=1, \dots, N} \text{sim}(\mathbf{y}_j^{\text{emb}}, \mathbf{x}_i^{\text{emb}}) \end{equation}

Finally, the F-measure score is computed by combining precision and recall (Equation~\ref{eq:f_measure}).

The metrics are calculated at the user story level but are averaged across the backlog to provide a broader perspective. This approach ensures that the evaluation considers variations in context and patterns across different backlogs.

In conclusion, the evaluation script provides a comprehensive framework for assessing the accuracy of extracted knowledge graph components. It not only enables a systematic comparison of different approaches but also establishes a foundation for future evaluations. 

\section{Experimental Setup}
\label{sec:setup}
To assess the accuracy of the \ac{USGT} module, two experiments were set up. The difference between the experiments lies on the LLM provider and its model chosen, the first is Llama 3 by Meta, an open-source, cost-free for research purposes and highly powerful model, and the second is GPT-4o mini, the most cost-efficient small model by OpenAI. These models were also chosen because Llama 3 does not support function calls, while GPT-4o mini does support this functionality, therefore demonstrating the versatility of the solution to work with different LLM providers. 

The dataset employed for experimentation is the same as in the experiment of Arulmohan et al. \cite{arulmohan2023extracting} and presented in Chapter~\ref{chap:foundation}. Which means, a cleaned version of the annotated dataset \cite{qualified-user-stories_2023}, representing 87\% of the complete version, to ensure the results of this thesis can be compared to their results.

For the LLM configuration, both models, Llama 3 and GPT-4o mini, were set to a zero temperature. This configuration parameter is necessary to control the stochasticity of the output, making it more consistent and predictable, since the output of the LLM is non-deterministic. 

To enable reproducibility and facilitate further experimentation with the \ac{USGT} module, the accompanying repository is organized as follows:

\begin{itemize}
    \item \textbf{\texttt{us\_graph\_transformer.py}}:  
    A Python file containing the \ac{USGT} module, including all its associated functions and components.

    \item \textbf{\texttt{pos\_baseline/}}:  
    A folder containing the annotated backlog dataset. This dataset serves as the ground truth for evaluation.

    \item \textbf{\texttt{extractor.py}}:  
    The automated script for running the \ac{USGT}. By default, it processes the backlog files from the \texttt{pos\_baseline/} folder but can be configured to extract data from alternative sources.

    \item \textbf{\texttt{extracted-user-stories/}}:  
    A folder to store the user stories extracted by the \texttt{extractor.py}. Each subfolder corresponds to a specific experiment, with results saved in JSON format.

    \item \textbf{\texttt{template.json}}:  A template JSON file demonstrating how the extracted user stories should be formatted for evaluation.

    \item \textbf{\texttt{evaluation.py}}:  The automated script to evaluate the results of an experiment by comparing the extracted user stories to the \texttt{pos\_baseline} (ground truth) dataset using the metrics defined on the previous section.

    \item \textbf{\texttt{evaluation/}}:  
    A folder to store evaluation results, including detailed comparison metrics for each experiment.
\end{itemize}

To replicate the experiments, readers can follow these steps:
\begin{enumerate}
    \item \textbf{Setup the environment:}  
    Clone the repository and install the required dependencies listed in \texttt{requirements.txt}.

    \item \textbf{Create a .env file:}  
    Create this file to store the LLM API key, if it is required by the LLM provider you chose, and the Neo4j connection parameters.
    
    \item \textbf{Prepare the experiment:} 
    In the \texttt{extractor.py} file, define the variables such as experiment name, and the LLM configuration. The two configurations explored in this thesis are pre-defined and can be uncommented.

    \item \textbf{Run the experiment:}
    Execute the \texttt{extractor.py} script. By default, the script processes data in \texttt{pos\_baseline/} folder. Modify the source in the script if using other datasets.

    \item \textbf{Evaluate the results:}  
    Compare the extracted user stories with the ground truth using the \texttt{evaluation.py} script. Evaluation metrics are saved in the \texttt{evaluation/} folder.
\end{enumerate}

The experiments involved configuring the LLMs and running the \texttt{extractor.py} script to apply the \ac{USGT} module to extract knowledge graphs from all the user stories within the 22 product backlogs, and then evaluating the results against the ground truth.

\section{Results}
\label{sec:results}
This section presents the outcomes of the experiments conducted to evaluate the performance of the \ac{USGT} module in extracting nodes according to the ontology defined. The results are analyzed to assess the accuracy, consistency, and adaptability of the module when working with different LLM providers. Specifically, the outputs of Llama 3 and GPT-4o mini are compared.

\begin{table}[h]
\centering
\setlength{\tabcolsep}{2pt} 
\renewcommand{\arraystretch}{1.2} 
\scriptsize 
\begin{tabular}{|c|c|c|c|c|c|c|c|c|}
\hline
\textbf{Backlog Name} & \multicolumn{2}{c|}{\textbf{Persona F-Measure}} & \multicolumn{2}{c|}{\textbf{Entity F-Measure}} & \multicolumn{2}{c|}{\textbf{Action F-Measure}} & \multicolumn{2}{c|}{\textbf{Benefit F-Measure}} \\
\cline{2-9}
 & GPT-4o-mini & Llama 3 & GPT-4o-mini & Llama 3 & GPT-4o-mini & Llama 3 & GPT-4o-mini & Llama 3 \\
\hline
g02 & 1.00 & 0.99 & 0.82 & 0.59 & 0.73 & 0.69 & 0.78 & 0.43 \\
g03 & 1.00 & 0.96 & 0.82 & 0.53 & 0.81 & 0.43 & 0.93 & 0.95 \\
g04 & 0.98 & 0.93 & 0.76 & 0.49 & 0.74 & 0.61 & 0.88 & 0.94 \\
g05 & 1.00 & 1.00 & 0.79 & 0.36 & 0.82 & 0.51 & 0.85 & 0.75 \\
g08 & 1.00 & 0.96 & 0.85 & 0.54 & 0.57 & 0.43 & 0.80 & 0.97 \\
g10 & 1.00 & 0.97 & 0.74 & 0.47 & 0.73 & 0.60 & 0.80 & 0.81 \\
g11 & 1.00 & 0.99 & 0.81 & 0.53 & 0.72 & 0.58 & 0.90 & 0.94 \\
g12 & 1.00 & 0.95 & 0.73 & 0.42 & 0.76 & 0.66 & 0.73 & 0.82 \\
g13 & 1.00 & 0.98 & 0.65 & 0.44 & 0.57 & 0.51 & 0.78 & 0.71 \\
g14 & 1.00 & 0.98 & 0.73 & 0.48 & 0.73 & 0.62 & 0.80 & 0.78 \\
g16 & 1.00 & 1.00 & 0.95 & 0.67 & 0.31 & 0.60 & 1.00 & 1.00 \\
g17 & 1.00 & 1.00 & 0.72 & 0.36 & 0.70 & 0.72 & 0.98 & 0.98 \\
g18 & 1.00 & 0.99 & 0.81 & 0.57 & 0.74 & 0.67 & 0.89 & 0.91 \\
g19 & 1.00 & 1.00 & 0.83 & 0.55 & 0.80 & 0.66 & 0.93 & 0.77 \\
g21 & 1.00 & 0.78 & 0.79 & 0.41 & 0.72 & 0.51 & 0.71 & 0.78 \\
g22 & 1.00 & 0.97 & 0.72 & 0.32 & 0.70 & 0.53 & 0.91 & 0.99 \\
g23 & 1.00 & 0.99 & 0.85 & 0.75 & 0.91 & 0.90 & 1.00 & 0.98 \\
g24 & 0.98 & 0.88 & 0.82 & 0.50 & 0.68 & 0.55 & 0.63 & 0.87 \\
g25 & 1.00 & 0.99 & 0.81 & 0.49 & 0.87 & 0.85 & 0.96 & 0.93 \\
g26 & 1.00 & 0.97 & 0.78 & 0.49 & 0.78 & 0.63 & 0.82 & 0.82 \\
g27 & 1.00 & 0.93 & 0.70 & 0.43 & 0.69 & 0.55 & 0.70 & 0.80 \\
g28 & 1.00 & 1.00 & 0.81 & 0.61 & 0.86 & 0.74 & 0.98 & 0.90 \\
\hline
\textbf{Average} & \textbf{1.00} & \textbf{0.96} & \textbf{0.79} & \textbf{0.50} & \textbf{0.73} & \textbf{0.62} & \textbf{0.85} & \textbf{0.85} \\
\hline
\end{tabular}
\caption{Strict Comparison of F-Measures for GPT-4o-mini and Llama 3.}
\label{tab:res_strict}
\end{table}

Based on the previously described evaluation methods, Table~\ref{tab:res_strict} presents results in the strict mode, Table~\ref{tab:res_inclusive} evaluates the inclusive mode, Table~\ref{tab:res_relax} evaluates the relaxed mode, and, lastly, Table~\ref{tab:bert} presents a comparison using BertScore.

In the strict mode comparison (Table~\ref{tab:res_strict}), GPT-4o-mini outperforms Llama 3 in extracting all node types except for \textit{benefit}, where Llama 3 shows a slightly higher performance. On average, Llama 3 struggles primarily with \textit{entity} and \textit{action} extraction in this mode. A detailed analysis of the data reveals that Llama 3 frequently mishandles noun qualifiers and verb complements, both of which are critical in this strict mode. Additionally, a significant performance gap is observed in backlog \textit{g02} for \textit{benefit} extraction. Upon closer inspection, many user stories in this backlog did not have a benefit to extract, yet Llama 3 either attempted to extract other parts from the story or produced hallucinations as the \textit{benefit} node.

\begin{table}
\centering
\setlength{\tabcolsep}{2pt} 
\renewcommand{\arraystretch}{1.2} 
\scriptsize 
\begin{tabular}{|c|c|c|c|c|c|c|c|c|}
\hline
\textbf{Backlog Name} & \multicolumn{2}{c|}{\textbf{Persona F-Measure}} & \multicolumn{2}{c|}{\textbf{Entity F-Measure}} & \multicolumn{2}{c|}{\textbf{Action F-Measure}} & \multicolumn{2}{c|}{\textbf{Benefit F-Measure}} \\
\cline{2-9}
 & GPT-4o-mini & Llama 3 & GPT-4o-mini & Llama 3 & GPT-4o-mini & Llama 3 & GPT-4o-mini & Llama 3 \\
\hline
g02 & 1.00 & 0.99 & 0.82 & 0.59 & 0.74 & 0.69 & 0.78 & 0.43 \\
g03 & 1.00 & 0.96 & 0.82 & 0.53 & 0.81 & 0.43 & 0.93 & 0.95 \\
g04 & 0.98 & 0.93 & 0.77 & 0.49 & 0.75 & 0.61 & 0.88 & 0.94 \\
g05 & 1.00 & 1.00 & 0.79 & 0.37 & 0.82 & 0.52 & 0.85 & 0.75 \\
g08 & 1.00 & 0.96 & 0.86 & 0.55 & 0.57 & 0.43 & 0.80 & 0.97 \\
g10 & 1.00 & 0.97 & 0.75 & 0.47 & 0.75 & 0.60 & 0.80 & 0.81 \\
g11 & 1.00 & 0.99 & 0.81 & 0.53 & 0.73 & 0.59 & 0.90 & 0.94 \\
g12 & 1.00 & 0.95 & 0.75 & 0.45 & 0.77 & 0.66 & 0.76 & 0.84 \\
g13 & 1.00 & 0.98 & 0.65 & 0.44 & 0.63 & 0.53 & 0.78 & 0.71 \\
g14 & 1.00 & 0.98 & 0.74 & 0.48 & 0.74 & 0.63 & 0.80 & 0.78 \\
g16 & 1.00 & 1.00 & 0.95 & 0.67 & 0.46 & 0.60 & 1.00 & 1.00 \\
g17 & 1.00 & 1.00 & 0.73 & 0.37 & 0.77 & 0.72 & 0.98 & 0.98 \\
g18 & 1.00 & 0.99 & 0.81 & 0.58 & 0.79 & 0.67 & 0.89 & 0.92 \\
g19 & 1.00 & 1.00 & 0.83 & 0.55 & 0.81 & 0.66 & 0.93 & 0.77 \\
g21 & 1.00 & 0.78 & 0.79 & 0.42 & 0.73 & 0.51 & 0.72 & 0.78 \\
g22 & 1.00 & 0.97 & 0.72 & 0.32 & 0.75 & 0.54 & 0.92 & 0.99 \\
g23 & 1.00 & 0.99 & 0.86 & 0.77 & 0.91 & 0.90 & 1.00 & 0.98 \\
g24 & 0.98 & 0.88 & 0.82 & 0.50 & 0.74 & 0.55 & 0.63 & 0.88 \\
g25 & 1.00 & 0.99 & 0.81 & 0.49 & 0.88 & 0.85 & 0.96 & 0.93 \\
g26 & 1.00 & 0.97 & 0.79 & 0.49 & 0.80 & 0.65 & 0.84 & 0.82 \\
g27 & 1.00 & 0.93 & 0.74 & 0.46 & 0.72 & 0.56 & 0.70 & 0.80 \\
g28 & 1.00 & 1.00 & 0.81 & 0.62 & 0.87 & 0.75 & 0.98 & 0.90 \\
\hline
\textbf{Average} & \textbf{1.00} & \textbf{0.96} & \textbf{0.79} & \textbf{0.51} & \textbf{0.75} & \textbf{0.62} & \textbf{0.86} & \textbf{0.86} \\
\hline
\end{tabular}
\caption{Inclusive Comparison of F-Measures for GPT-4o-mini and Llama 3.}
\label{tab:res_inclusive}
\end{table}

Moving to the inclusive and (Table~\ref{tab:res_inclusive}) relaxed (Table~\ref{tab:res_relax}) modes, both models show improved scores across all categories as expected. However, GPT-4o-mini continues to lead with a more consistent performance across all categories, maintaining its edge in benefit extraction in the inclusive mode.

\begin{table}
\centering
\setlength{\tabcolsep}{4pt} 
\renewcommand{\arraystretch}{1.2} 
\scriptsize 
\begin{tabular}{|c|c|c|c|c|c|c|}
\hline
\textbf{Backlog Name} & \multicolumn{2}{c|}{\textbf{Persona F-Measure}} & \multicolumn{2}{c|}{\textbf{Entity F-Measure}} & \multicolumn{2}{c|}{\textbf{Action F-Measure}} \\
\cline{2-7}
 & GPT-4o-mini & Llama 3 & GPT-4o-mini & Llama 3 & GPT-4o-mini & Llama 3 \\
\hline
g02 & 1.00 & 0.99 & 0.89 & 0.64 & 0.79 & 0.73 \\
g03 & 1.00 & 0.96 & 0.86 & 0.60 & 0.85 & 0.53 \\
g04 & 0.98 & 0.93 & 0.84 & 0.55 & 0.80 & 0.64 \\
g05 & 1.00 & 1.00 & 0.86 & 0.40 & 0.85 & 0.59 \\
g08 & 1.00 & 0.96 & 0.89 & 0.56 & 0.62 & 0.49 \\
g10 & 1.00 & 0.97 & 0.82 & 0.53 & 0.81 & 0.68 \\
g11 & 1.00 & 0.99 & 0.87 & 0.58 & 0.77 & 0.64 \\
g12 & 1.00 & 0.95 & 0.80 & 0.48 & 0.84 & 0.74 \\
g13 & 1.00 & 0.98 & 0.73 & 0.49 & 0.67 & 0.60 \\
g14 & 1.00 & 0.98 & 0.82 & 0.55 & 0.76 & 0.68 \\
g16 & 1.00 & 1.00 & 0.95 & 0.67 & 0.31 & 0.60 \\
g17 & 1.00 & 1.00 & 0.81 & 0.41 & 0.72 & 0.72 \\
g18 & 1.00 & 0.99 & 0.85 & 0.60 & 0.80 & 0.71 \\
g19 & 1.00 & 1.00 & 0.88 & 0.58 & 0.82 & 0.69 \\
g21 & 1.00 & 0.78 & 0.84 & 0.47 & 0.79 & 0.58 \\
g22 & 1.00 & 0.98 & 0.83 & 0.36 & 0.73 & 0.65 \\
g23 & 1.00 & 0.99 & 0.89 & 0.79 & 0.96 & 0.92 \\
g24 & 0.98 & 0.88 & 0.86 & 0.52 & 0.74 & 0.61 \\
g25 & 1.00 & 0.99 & 0.92 & 0.58 & 0.87 & 0.85 \\
g26 & 1.00 & 0.98 & 0.84 & 0.53 & 0.85 & 0.73 \\
g27 & 1.00 & 0.93 & 0.76 & 0.47 & 0.76 & 0.61 \\
g28 & 1.00 & 1.00 & 0.88 & 0.63 & 0.89 & 0.76 \\
\hline
\textbf{Average} & \textbf{1.00} & \textbf{0.97} & \textbf{0.85} & \textbf{0.54} & \textbf{0.77} & \textbf{0.67} \\
\hline
\end{tabular}
\caption{Relaxed Comparison of F-Measures for GPT-4o-mini and Llama 3.}
\label{tab:res_relax}
\end{table}

\begin{table}
\centering
\setlength{\tabcolsep}{2pt} 
\renewcommand{\arraystretch}{1.2} 
\scriptsize 
\begin{tabular}{|c|c|c|c|c|c|c|c|c|}
\hline
\textbf{Backlog Name} & \multicolumn{2}{c|}{\textbf{Persona F-Measure}} & \multicolumn{2}{c|}{\textbf{Entity F-Measure}} & \multicolumn{2}{c|}{\textbf{Action F-Measure}} & \multicolumn{2}{c|}{\textbf{Benefit F-Measure}} \\
\cline{2-9}
 & GPT-4o-mini & Llama 3 & GPT-4o-mini & Llama 3 & GPT-4o-mini & Llama 3 & GPT-4o-mini & Llama 3 \\
\hline
g02 & 1.00 & 0.99 & 0.95 & 0.87 & 0.85 & 0.79 & 0.92 & 0.55 \\
g03 & 1.00 & 0.99 & 0.93 & 0.79 & 0.88 & 0.74 & 0.98 & 1.00 \\
g04 & 1.00 & 0.99 & 0.94 & 0.84 & 0.89 & 0.82 & 0.98 & 1.00 \\
g05 & 1.00 & 1.00 & 0.94 & 0.75 & 0.90 & 0.78 & 0.96 & 1.00 \\
g08 & 1.00 & 0.97 & 0.94 & 0.77 & 0.83 & 0.82 & 0.94 & 1.00 \\
g10 & 1.00 & 0.98 & 0.92 & 0.85 & 0.88 & 0.79 & 0.97 & 1.00 \\
g11 & 1.00 & 0.99 & 0.93 & 0.84 & 0.88 & 0.72 & 0.97 & 1.00 \\
g12 & 1.00 & 0.97 & 0.93 & 0.86 & 0.86 & 0.84 & 0.98 & 0.96 \\
g13 & 1.00 & 0.99 & 0.88 & 0.86 & 0.81 & 0.74 & 1.00 & 1.00 \\
g14 & 1.00 & 0.98 & 0.91 & 0.86 & 0.87 & 0.79 & 0.97 & 1.00 \\
g16 & 1.00 & 1.00 & 0.93 & 0.87 & 0.72 & 0.67 & 1.00 & 1.00 \\
g17 & 1.00 & 1.00 & 0.91 & 0.79 & 0.86 & 0.81 & 1.00 & 0.98 \\
g18 & 1.00 & 1.00 & 0.93 & 0.83 & 0.89 & 0.79 & 0.97 & 0.93 \\
g19 & 1.00 & 1.00 & 0.93 & 0.87 & 0.87 & 0.78 & 0.96 & 0.78 \\
g21 & 1.00 & 0.98 & 0.91 & 0.81 & 0.84 & 0.77 & 0.96 & 1.00 \\
g22 & 1.00 & 0.99 & 0.94 & 0.81 & 0.82 & 0.76 & 0.99 & 1.00 \\
g23 & 1.00 & 0.99 & 0.97 & 0.96 & 0.97 & 0.93 & 1.00 & 0.98 \\
g24 & 1.00 & 0.99 & 0.92 & 0.78 & 0.87 & 0.81 & 0.96 & 1.00 \\
g25 & 1.00 & 1.00 & 0.95 & 0.86 & 0.92 & 0.90 & 0.96 & 0.93 \\
g26 & 1.00 & 0.99 & 0.92 & 0.87 & 0.92 & 0.85 & 0.99 & 1.00 \\
g27 & 1.00 & 0.99 & 0.91 & 0.79 & 0.85 & 0.78 & 0.84 & 0.83 \\
g28 & 1.00 & 1.00 & 0.94 & 0.86 & 0.93 & 0.85 & 0.98 & 0.88 \\
\hline
\textbf{Average} & \textbf{1.00} & \textbf{0.99} & \textbf{0.93} & \textbf{0.84} & \textbf{0.87} & \textbf{0.80} & \textbf{0.97} & \textbf{0.95} \\
\hline
\end{tabular}
\caption{BertScore Comparison of F-Measures for GPT-4o-mini and Llama 3.}
\label{tab:bert}
\end{table}

The BertScore (Table~\ref{tab:bert}) comparison mode brings the semantic alignment perspective, and in this case both models show a good performance. 

\begin{figure}
  \centering
  \includegraphics[width=0.5\textwidth]{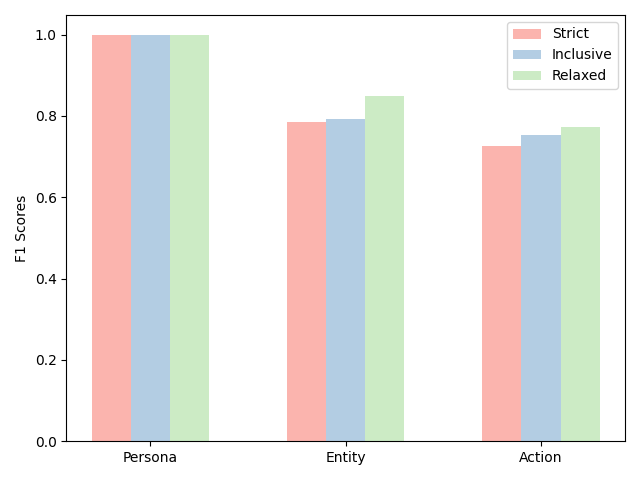}
  \caption{Average comparison of node types extraction using GPT-4o-mini model.}
  \label{fig:graph_gpt}
\end{figure}

\begin{figure}
  \centering
  \includegraphics[width=0.5\textwidth]{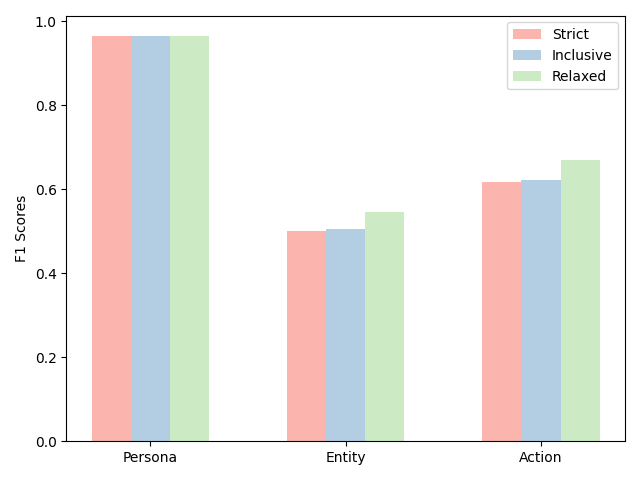}
  \caption{Average comparison of node types extraction using LLama3.}
  \label{fig:graph_ollama}
\end{figure}

Overall, GPT-4o-mini (Figure~\ref{fig:graph_gpt}) demonstrates superior performance across all evaluated categories, with consistently higher F-Measure scores. Even though it doesn't have a perfect F-measure when comparing exact string matching, it proves capable in capturing semantic alignment, particularly in the \textit{benefit} node, which poses a great challenge since it has many elements to be extracted. Llama 3 (Figure~\ref{fig:graph_ollama}), while competitive, exhibits greater variability.

\section{Discussion}
\label{sec:discussion}
This study aimed to investigate the effectiveness of the \ac{USGT} module in extracting knowledge graphs from user stories. Specifically, to compare the performance of the proposed approach to existing methods, explore the feasibility of a generalized approach across different LLMs, and assess the potential for a fully automated solution. In this section, the results will be discussed, a comparison to the previous experiment is presented and the research questions are answered.

\begin{quote}
\textit{RQ1.How does the accuracy of nodes and relationships extraction of the proposed solution compare to the existing Large Language Model-based method~\cite{arulmohan2023extracting}?}
\end{quote}

To answer the first research question it is necessary to perform an equivalent experiment to the one presented by Arulmohan et al.~\cite{arulmohan2023extracting}. The existing solution applied GPT-3.5 model by OpenAI to extract the nodes and relationships of user stories. However, as of November 2024, this model had been deprecated, and is no longer available for use in the experiment proposed by this thesis. 

Fortunately, the results from the same research group as in~\cite{arulmohan2023extracting} can be found on their online repository~\cite{sebresults}. These results applied the same techniques described in the published paper~\cite{arulmohan2023extracting}, but with different GPT models, including GPT-4-turbo. This access enabled a comparison between the solution proposed in this thesis and the existing approach using GPT-4-0125-preview (also referred to as GPT-4-turbo), as well as the \ac{CRF} method, as summarized in Table~\ref{tab:benchmark}. The experiment was conducted using the same dataset and a zero temperature configuration on the LLM.

\begin{table}[!ht]
\centering
\setlength{\tabcolsep}{4pt} 
\renewcommand{\arraystretch}{1.2} 
\scriptsize 
\begin{tabular}{|c|ccc|ccc|ccc|}
\hline
\textbf{Backlog Name} & \multicolumn{3}{c|}{\textbf{Benchmark GPT-4-turbo}} & \multicolumn{3}{c|}{\textbf{GPT-4-turbo}} & \multicolumn{3}{c|}{\textbf{CRF}} \\
\cline{2-10}
 & \textbf{P} & \textbf{E} & \textbf{A} & \textbf{P} & \textbf{E} & \textbf{A} & \textbf{P} & \textbf{E} & \textbf{A} \\
\hline
g02 & 1.00 & 0.67 & 0.66 & 1.00 & 0.84 & 0.76 & 1.00 & 0.77 & 0.83 \\
g03 & 1.00 & 0.76 & 0.81 & 0.84 & 0.85 & 0.89 & 1.00 & 0.82 & 0.90 \\
g04 & 0.94 & 0.69 & 0.71 & 1.00 & 0.80 & 0.82 & 1.00 & 0.77 & 0.83 \\
g05 & 1.00 & 0.69 & 0.66 & 1.00 & 0.83 & 0.82 & 1.00 & 0.85 & 0.86 \\
g08 & 1.00 & 0.73 & 0.74 & 1.00 & 0.84 & 0.64 & 1.00 & 0.97 & 0.68 \\
g10 & 0.98 & 0.62 & 0.71 & 1.00 & 0.73 & 0.75 & 1.00 & 0.81 & 0.76 \\
g11 & 1.00 & 0.69 & 0.79 & 1.00 & 0.81 & 0.78 & 1.00 & 0.84 & 0.81 \\
g12 & 1.00 & 0.64 & 0.69 & 1.00 & 0.73 & 0.75 & 1.00 & 0.72 & 0.73 \\
g13 & 1.00 & 0.56 & 0.60 & 1.00 & 0.70 & 0.65 & 1.00 & 0.63 & 0.60 \\
g14 & 1.00 & 0.63 & 0.71 & 1.00 & 0.76 & 0.76 & 1.00 & 0.76 & 0.84 \\
g17 & 1.00 & 0.68 & 0.78 & 1.00 & 0.78 & 0.83 & 1.00 & 0.83 & 0.80 \\
g18 & 1.00 & 0.73 & 0.69 & 1.00 & 0.82 & 0.78 & 1.00 & 0.73 & 0.84 \\
g19 & 1.00 & 0.79 & 0.74 & 1.00 & 0.85 & 0.86 & 1.00 & 0.88 & 0.88 \\
g21 & 0.84 & 0.65 & 0.67 & 1.00 & 0.73 & 0.73 & 1.00 & 0.82 & 0.77 \\
g22 & 1.00 & 0.59 & 0.78 & 0.99 & 0.74 & 0.80 & 1.00 & 0.78 & 0.88 \\
g23 & 1.00 & 0.81 & 0.87 & 1.00 & 0.86 & 0.89 & 1.00 & 0.88 & 0.97 \\
g24 & 0.94 & 0.64 & 0.65 & 1.00 & 0.83 & 0.73 & 1.00 & 0.82 & 0.79 \\
g25 & 1.00 & 0.79 & 0.84 & 1.00 & 0.85 & 0.91 & 1.00 & 0.91 & 0.94 \\
g26 & 1.00 & 0.64 & 0.71 & 1.00 & 0.77 & 0.79 & 1.00 & 0.81 & 0.91 \\
g27 & 0.98 & 0.58 & 0.67 & 0.99 & 0.72 & 0.73 & 1.00 & 0.76 & 0.82 \\
g28 & 1.00 & 0.81 & 0.84 & 1.00 & 0.87 & 0.86 & 1.00 & 0.75 & 0.88 \\
\hline
\textbf{Average} & \textbf{0.98} & \textbf{0.69} & \textbf{0.73} & \textbf{0.99} & \textbf{0.80} & \textbf{0.79} & \textbf{1.00} & \textbf{0.81} & \textbf{0.82} \\
\hline
\end{tabular}
\caption{Strict Comparison of F-Measures for GPT-4-turbo from Benchmark, GPT-4-turbo using \ac{USGT} module, and \ac{CRF}.}
\label{tab:benchmark}
\end{table}

When comparing results, the \ac{USGT} module demonstrated superior performance on average across all node types. Notably, entity extraction improved by 16\% on average compared to the benchmark. However, there were exceptions, such as persona extraction for backlog g03, which performed poorly compared to the benchmark. This discrepancy arose from the complexity of the persona in this backlog, which included multiple elements such as role and specific context or function (e.g., Planning Staff Member).

The differences in results are partly attributable to internal changes in the LLMs over time, as the experiments were conducted at different moments. However, a significant factor was the variation in prompt designs. While both experiments followed OpenAI's prompt guidelines~\cite{openai_prompt}, the implementation of the guideline's strategies differed substantially, the wording, phrase construction and instructions to the LLM are different and this is a common challenge when working with this technology.

The earlier study also employed \ac{CRF}, a tailored machine-learning approach trained on 20\% of the dataset. This method outperformed all GPT models and the Visual Narrator technique (details in Chapter~\ref{chap:foundation}). When comparing \ac{CRF} to GPT-4-turbo using \ac{USGT}, \ac{CRF} still leads in performance; however, the gap has narrowed under the same evaluation metrics, making LLM-based solutions a competitive alternative. To determine the most assertive approach, it is necessary to consider specific use case and context. For tasks that require constant updates, flexibility, or are part of larger, dynamic systems, LLMs may offer advantages. On the other hand, for applications where efficiency, quick deployment, and low computational cost are essential, \ac{CRF} might be a better choice. 

The previous work also mentioned the challenges when depending on a specific hosting provider, including technical changes affecting API interaction and data processing, as well as pricing, and model availability. 

In contrast, this thesis proposes a versatile, model-agnostic solution, as demonstrated by experiments using models from two different providers. This approach addresses prior limitations by enabling the selection of models and providers based on specific contexts and constraints. For instance, in a budget-constrained scenario, an open-source LLM provider could be used, while in experimental contexts, the solution facilitates comparison across different models and providers.

This study aimed to investigate the effectiveness of the \ac{USGT} module in extracting knowledge graphs from user stories. Specifically, to compare the performance of the proposed approach to existing methods, explore the feasibility of a generalized approach across different LLMs, and assess the potential for a fully automated solution. 

The results previously reported show that the solution is able to successfully extract nodes and relationships from user stories according to the specific ontology proposed, however the adherence to the ontology, and therefore, the quality of the results depends on the chosen large language model. 

\begin{quote}
\textit{RQ2. How can a knowledge graph be automatically generated from user stories using large language models without being dependent on a specific provider?}
\end{quote}

The second research question directly addresses the versatility of the solution. By using the LangChain framework, a unified module is created, that can integrate with various LLM providers. This abstraction layer allows the module to easily configure and use a new language model, without modifying the core logic of it. This flexibility not only accelerates development but also enables us to explore the strengths and weaknesses of different models to optimize performance.

\begin{quote}
\textit{RQ3. What strategies can be employed to build a fully automated solution to extract knowledge graph from user stories?}
\end{quote}

Previous research~\cite{arulmohan2023extracting} demonstrated that large language models (LLMs), despite certain limitations in accuracy, have the capability to extract the core components of a knowledge graph—namely, nodes and relationships—from user stories. These extracted components hold significant potential as tools to enhance the requirements engineering life cycle. However, the prior work primarily focused on the extraction process and did not address the practical implementation of a fully realized knowledge graph. Traditional methods for constructing knowledge graphs often involve substantial complexity and effort. To bridge this gap, this thesis proposes a comprehensive solution that not only extracts the knowledge graph components but also implements them in an automated and streamlined manner.

To achieve a complete degree of automation, the solution proposed relied on LangChain in these aspects:
\begin{itemize}
    \item Defining a prompt template: the prompt template is defined in advance and has a placeholder that dynamically adds the input user story to it, without the necessity of manual adjustments each time a new user story is received.

    \item Interacting with API via chains: through \ac{LCEL} it is possible to define chains to standardize the API interaction. This eliminates the need for manual API calls and data formatting at each step, automating the process of feeding inputs to LLMs and processing the outputs.

    \item Using pre-built modules: LangChain offers a rich set of ready-to-use modules that facilitate various tasks such as configuring LLMs, performing data transformations, and interacting with external databases (e.g., Neo4j). These modules help streamline the entire process, from model configuration to data storage, and therefore, significantly reduce implementation complexity.

\end{itemize}

This approach not only automates the extraction and storage of knowledge graphs but also enables the system to be easily extended or modified by simply adjusting the relevant LangChain components, making it highly adaptable for future use cases or updates.

\chapter{State of the Art}
\label{chap:state}
While user stories are a popular tool in Requirements Engineering (RE) for capturing user needs, extracting and structuring the information they contain can be a challenge. This chapter explores how recent advancements in two key areas - Knowledge Graphs (KGs) and Large Language Models (LLMs) - can offer new solutions for RE, particularly when focused on user stories, as well as how LangChain is being used in research. We will delve into existing research on these three pillars and their potential intersections, examining how they can be leveraged to enhance the clarity, structure, and overall utility of modeled user requirements.

\section{Modeling User Stories using NLP}
Modeling user stories can be understood as part of domain modeling \cite{broy2013domain}, which involves identifying essential business and application entities and their relationships. \ac{NLP} is the most common technique to model user stories. The study by Raharjana et al. \cite{raharjana2021user} identifies eight approaches centered on modeling, which can be broadly categorized into three groups based on their methodologies and objectives \cite{mosser2022modelling}:

\begin{itemize}
    \item \textbf{Conceptual Modelling with Visual Narrator Tool:} 
    Three of the identified approaches utilize the Visual Narrator tool \cite{robeer2016automated}, developed by the Requirements Engineering Lab at Utrecht University, to extract conceptual models, a textual template to define a structured representation of user stories, from backlogs. This tool, which creates ontology-like structures, is used differently to analyze user stories. The first approach employs \ac{POS} tagging to parse the user stories and construct the conceptual models Lucassen et al. \cite{lucassen2016improving}. Despite its utility, this method suffers from low precision in accurately identifying parts of speech, which can lead to errors in the resulting models. Similarly, the second approach by Lucassen et al. \cite{lucassen2017extracting} relies on \ac{POS} tagging but encounters additional difficulties with the correct identification of compound terms, impacting the overall results. The third approach by Dalpiaz et al. \cite{dalpiaz2019detecting} diverges by utilizing a vector space model in conjunction with manual data tagging. 

    \item \textbf{Generation of UML Artifacts:}
    Another subset of three approaches focuses on transforming user stories into various UML artifacts using \ac{POS} tagging technique. This includes the creation of class diagrams by Nasiri et al. \cite{nasiri2021towards}, sequence diagrams by Elallaoui et al. \cite{elallaoui2015automatic}, and use case diagrams \cite{elallaoui2018automatic}, although useful for generating a basic overview of system interactions, this technique fails in handling the nuances of more sophisticated systems, limiting its applicability in complex environments.

    \item \textbf{Ontology-Based Approaches:}
    The final two approaches adopt an ontology-based perspective \cite{genaid2012connecting} \cite{athiththan2018ontology}. The first method by Genaid et al. \cite{genaid2012connecting} involves constructing an ontology directly from the source code and using \ac{NER} in user stories with specific code locations. This approach automates traceability, ensuring that user stories are accurately linked to their corresponding code segments, however, it presents a low precision. The second method by Athiththan et al. \cite{athiththan2018ontology} models the backlog itself as an ontology using \ac{POS} tagging, which is then used to generate boilerplate code, thus speeding up the development process. Despite their innovative use of ontologies, these approaches do not fully utilize the graph structure inherent in ontologies.
\end{itemize}

Despite numerous studies using NLP, empirical evaluations demonstrating concrete effectiveness are scarce \cite{zhao2021natural}. In addition, these approaches are limited by their precision and recall \cite{raharjana2021user}, therefore this thesis aims to improve these metrics.

\subsection{Ontologies}
A key aspect of domain modeling is ontology, a formal representation of domain modeling. A user story modeling ontology was proposed by Mancuso et al.~\cite{mancuso2023approach} which created a domain-specific modeling language and integrated it into the modeling tool AOAME that resulted in a visual user story. Ladeinde et al.~\cite{ladeinde2023extracting} also proposed the use of \acp{KG} to model user stories, by applying \ac{NLP} techniques, and its ontology is based on role, goal, and benefit.

While the ontologies proposed by Mancuso et al. and Ladeinde et al. offer valuable approaches, Arulmohan et al.'s ontology is considered a more comprehensive and flexible framework for this research, because the first~\cite{mancuso2023approach} doesn't have general node types which difficult the querying, while the second~\cite{ladeinde2023extracting} extracts only three node types and doesn't standardize the relationships types.

\section{Large Language Models and Knowledge Graphs}
Large Language Models (LLMs) have shown potential for building knowledge graphs (KGs) due to their natural language understanding capabilities. They excel at understanding natural language, making them well-suited for two main tasks required for the construction of a \ac{KG}: concept extraction and relation extraction \cite{khorashadizadeh2024research}. Concept extraction involves identifying key ideas and topics within the text. Relation extraction focuses on how these concepts are connected. To improve this process, techniques like \ac{NER} and additional context should be added to the LLM.

Previous work relied on prompt engineering to extract entities, such as PromptNER by Ashok et al. \cite{ashok2023promptner} that proposes a solution based on four vital components: an LLM, a prompt to define and restrict the set of entity types, a concise collection of domain-specific examples, and instructions to specify the intended output format.

An alternative approach by Khorashadizadeh et al. \cite{khorashadizadeh2024research} is based on instruction-tuning, in which the language models receive natural language instructions to guide the model's response. In this direction, Zhou et al. \cite{zhou2023universalner} presented a target distillation approach with mission-focused instruction tuning.

While prompt engineering and instruction tuning offer advancements, they struggle with user stories due to limited ontology awareness. These approaches propose a generic entity extraction but lack the ability to understand the specific structure and constraints defined within a user story ontology. This leads to inaccurate or incomplete knowledge graphs, as extracted entities might not correspond to the ontology's specifications. Additionally, current LLM-based methods often lack scalability, hindering their transition to production-ready systems. The approach proposed in this thesis is tailored to capture user story nuances and benefits from the usage of a robust framework that supports LLM systems scalability.

\section{Large Language Models and Requirements Engineering}
There is a wide range of applications of LLMs in the domain of \ac{RE}~\cite{hou2023large}: anaphoric ambiguity treatment, requirements classification, coreference detection, requirements elicitation, and software traceability.

Research by White et al. \cite{white2024chatgpt} proposes prompt patterns that leverage LLMs, such as ChatGPT, to facilitate the elicitation and identification of missing requirements [2]. This approach can help capture user needs more comprehensively during the initial stages of software development.

Endres et al. \cite{endres2023formalizing} explore the use of LLMs to formalize requirements from natural language intent. This holds promise for streamlining the transition from user stories to formal specifications, improving clarity, and reducing ambiguity.

Arulmohan et al. \cite{arulmohan2023extracting} investigated GPT-3.5's potential for extracting domain models from user stories. While they proposed a framework for transforming these concepts into a comprehensive domain representation, they did not address the specific challenge of knowledge graph creation.

Subsequently, Bragilovski et al. \cite{bragilovski2024deriving} compared human, rule-based, GPT, and ML-based approaches for deriving domain models. Their findings highlighted that human performance still surpasses AI-based methods, indicating a need for further advancements in AI techniques.

Cheng et al.~\cite{cheng2024generative} explored the application of generative AI in requirements engineering, noting the widespread use of GPT models in this field. This thesis solution further contributes to this domain by offering a versatile model solution that expands upon existing research.

There remains a notable absence of research on employing LLMs for requirements engineering and design purposes. The literature review by Hou et al.~\cite{hou2023large} spanning 2017-2024 demonstrated that only 3.9\% of the research in the application of LLMs in software engineering was focused on requirements engineering. Additionally, Fan et al.\cite{fan2023large} highlight that practitioners are hesitant to rely on LLMs for higher-level design goals. This hesitancy underscores the need for further research to build trust and demonstrate the value proposition of LLMs in RE.

\section{LangChain Applications}
While LangChain has proven its value in various applications, its direct application to user story modeling and knowledge graph generation within the realm of requirements engineering remains relatively unexplored. A comprehensive literature review revealed a lack of research specifically addressing the integration of LangChain into these domains. This scarcity presents a promising path for original research and innovation.

 LangChain has been used to build sophisticated question-answering systems that can provide informative and accurate responses to user queries. Jeong's work \cite{jeong2024study} developed an advanced \textit{\ac{RAG}} system based on graph technology using LangChain, which resulted in increasing accuracy and relevance of the responses provided by the system to the users.

 Another application of this framework is in the integration of chatbots and virtual assistants. Singh et al. \cite{singh2024revolutionizing} explored the application of \ac{LLM}s through LangChain to introduce MindGuide, a chatbot to assist individuals looking for guidance regarding mental health. They incorporated features such as ChatPrompt Template and LLMChain to rapidly prototype and streamline the orchestration of the \ac{LLM} in their application.

However, to fully leverage LangChain's potential, it is essential to extend its application to the critical domain of requirements engineering. By exploring its utility in user story modeling and knowledge graph generation, this research aims to contribute to the advancement of the field and provide novel solutions to existing challenges.

\chapter{Conclusion}
\label{chap:conclusion}
In this chapter, we conclude the work presented in this thesis. Section~\ref{sec:summary} restates the research objectives and summarizes the key findings. Section~\ref{sec:contribution} highlights the contributions and the impact of the results. Section~\ref{sec:limitation} discusses the limitations of the study, and finally, Section~\ref{sec:future} explores potential paths for future research.

\section{Summary}
\label{sec:summary}
This thesis, motivated to improve the requirements engineering life-cycle,  aimed to investigate the effectiveness of the USGT module in extracting knowledge graphs from user stories. Specifically, we sought to compare the performance of our approach to existing methods, explore the feasibility of a generalized approach across different LLMs, and assess the potential for a fully automated solution.

The potential of the LangChain framework was explored to simplify the development of LLM-based solutions for knowledge graph generation. And, to facilitate reproducibility and comparison with existing methods, a reusable evaluation script was developed to assess the performance of the USGT module against a predefined ground truth, which can be used in the future for evaluation of new methods.

Through a comprehensive evaluation, the USGT module demonstrated superior performance compared to the existing LLM-based method, however still not better than a pre-trained method, such as \ac{CRF} particularly in terms of F-measure. The flexibility of the LangChain framework was also highlighted, enabling the seamless integration of different LLMs and facilitating the automation of the extraction process, therefore addressing a previous limitation.

\section{Contributions}
\label{sec:contribution}
This thesis makes significant contributions to the field of automated knowledge graph generation and its application within requirements engineering. First, it introduces the \ac{USGT} Python module, a novel tool designed to extract nodes and relationships from user stories and transform them into a knowledge graph format. This module builds on existing methodologies, particularly benefiting from the strengths of \acp{LLM} and the LangChain framework, to provide an automated approach for knowledge graph creation.

Additionally, this research proposes a practical solution tailored to the requirements engineering domain, offering practitioners a method to integrate the \ac{USGT} module to automate the generation of knowledge graphs and store it in a graph database.

A further contribution lies in the development of a robust evaluation script, which extends the evaluation criteria established by Arulmohan et al.~\cite{arulmohan2023extracting}. This script introduces a new evaluation metric, BERTScore, to assess semantic similarity between extracted and ground truth components and expands the evaluation to include an additional node type, the benefit node. The evaluation script not only supports reproducibility of the experiments presented in this study but also serves as a foundation for future research and experimentation in the field.

Moreover, this thesis conducts a comprehensive evaluation of the \ac{USGT} module's performance. A comparative assessment with prior work highlights improvements and areas for further refinement, demonstrating the value of the proposed approach.

The research also contributes to the exploration of the development of LLM-based applications, specifically in the context of knowledge graph construction. It showcases the potential of the LangChain framework in simplifying the integration of LLMs into workflows.

Through these contributions, the thesis aims to enhance the requirements engineering life-cycle by promoting transparency among stakeholders and development teams. By maintaining a manageable product backlog stored in a knowledge graph format, it can enable the identification and analysis of dependencies between requirements, supporting continuous improvement and fostering a more assertive and structured software development process.

\section{Limitations and Threats to Validity}
\label{sec:limitation}
While the proposed USGT module has shown promising results, it is important to acknowledge its limitations. The limitations categorized based on their impact on the different types of threats to validity, including construct, internal, external, and conclusion validity. 

One limitation concerns the evaluation against the ground truth. Although the dataset was annotated through a rigorous process, the exact annotation of node elements remains somewhat subjective. Different interpretations by scientists, requirements analysts, or engineers could lead to variations in the annotations, which may affect the accuracy of the evaluation. This subjectivity poses a threat to construct validity, as the evaluation might not consistently reflect the true quality or correctness of the extracted knowledge graph components.

Another limitation is the non-deterministic nature of LLM outputs. Even with model parameters such as temperature set to zero, responses from the LLM may vary across executions. This variability impacts internal validity, as it can lead to inconsistent experimental results, making it difficult to attribute outcomes solely to the factors being tested. Moreover, it affects external validity, as the same approach might yield inconsistent results when applied in different settings or with alternative datasets. 

Although the USGT module allows experimentation with LLMs from different providers, these models may exhibit biases or lack sufficient domain understanding. This limitation poses a threat to external validity, as the findings may not generalize well to diverse domains or user story formats. 

The flexibility of the module to support multiple LLM providers is made possible through LangChain, which simplifies implementation complexity. However, this reliance introduces a dependency on the framework itself, which could become problematic if LangChain undergoes significant changes, becomes obsolete, or imposes constraints that are incompatible with future requirements.

The definition of the prompt plays a critical role in the performance of the model. While the prompt is a key factor, its formulation remains highly subjective. Even though guidelines exist~\cite{openai_prompt}, they are constantly changing, and it is difficult to determine with certainty whether a prompt is \enquote{right} or \enquote{wrong} which brings an element of uncertainty into the process. This introduces a threat to construct validity, as prompt design choices might inadvertently influence what is being measured. Additionally, the reliance on subjective prompt formulations may also pose a threat to internal validity. 

Finally, while the proposed solution successfully automates the extraction and storage of knowledge graphs in Neo4j, the study does not evaluate the degree of automation in detail or assess the usability of the solution from a user perspective. This limitation impacts conclusion validity, as it leaves questions about the practicality and effectiveness of the solution for end-users unanswered. 

The limitations outlined in this study may impact the interpretation and generalizability of the results. However, acknowledging these limitations provides a transparent evaluation of the work and highlights opportunities for future research, which will be discussed in the following section.

\section{Future Avenues}
\label{sec:future}
This section presents potential paths for future research based on the findings and limitations of this study.

As presented in this study and in the state of the art, there are many methods and tools to extract the core components of a knowledge graph-nodes and relationships. These include methods such as the the Visual Narrator Tool, Conditional Random Fields, ChatGPT, and the \ac{USGT} module. While these methods differ in performance, they offer unique advantages depending on the specific context, such as implementation complexity, reliability of results, dataset size, available resources, and long-term stability. Future research could involve a comprehensive comparative analysis of these methods, accompanied by a practical guide to help practitioners choose the most suitable approach based on their specific requirements.

Focusing on \acp{LLM} methods for knowledge graph extraction, a recent work by Garbas et al.~\cite{garbas2024transformerranker} proposes a new tool, TransformerRanker, designed to evaluate language models that use a transformer architecture for classification tasks. This approach suggests a promising direction for future research: integrating the language model versatility of \ac{USGT} module with a component to identify the best language model to extract nodes and relationships from a backlog dataset. 

To build trust among requirements analysts and engineers who could benefit from the proposed solution, future research could evaluate the degree of automation achieved by the \ac{USGT} module. This includes quantifying the extent of automation across the entire solution and conducting usability testing to assess its practicality for end-users. Insights gained from such evaluations could inform improvements in user experience and help bridge the gap between academic research and real-world applications.

\ifdefined\ARXIV

\else
  \bibliographystyle{plain}
  \bibliography{references}
\fi

\end{document}